\documentstyle[a4,english,twoside,12pt]{article}

\newcommand{\resetcounter}{\setcounter{equation}{0}}     % set counter to zero

\begin{document}
\thispagestyle{empty}
%\rightline{IASSNS-HEP-98-59}
\rightline{HUB-EP-98/51}
\rightline{hep-th/9808159}
\vspace{2truecm}
\centerline{\bf \LARGE $N=1$ Heterotic/F-Theory Duality}

\vspace{1.5truecm}
\centerline{ 
Bj\"orn Andreas}\ , \ 
%Gottfried Curio$^\dagger$\footnote{curio@ias.edu}\ and \
%Dieter L\"ust$^*$\footnote{luest@physik.hu-berlin.de}}

\vspace{.2truecm}
{\em 
\centerline{Humboldt-Universit\"at, Institut f\"ur Physik,
D-10115 Berlin, Germany}}

%\vspace{.4truecm}
%{\em
%\centerline{$^\dagger$School of Natural Sciences, Institute for Advanced 
%Stud, Princeton, NJ 08540}}

%\vspace{1.0truecm}
%%%%%%%%%%%%%%%%%%%%%%%%%%%%%%%%%%%%%%%%%%%%%%%%%%%%%%%%%
%\vspace{.4truecm}
\begin{abstract}
We review aspects of $N=1$ duality between the heterotic string and F-theory. 
After a description of string duality intended for the non-specialist the 
framework and the constraints for heterotic/F-theory compactifications are 
presented. The computations of the necessary Calabi-Yau manifold
and vector bundle data, involving characteristic classes and bundle moduli,
are given in detail. The matching of the spectrum of chiral multiplets 
and of the number of heterotic five-branes respectively F-theory three-branes, 
needed for anomaly cancellation in four-dimensional vacua, 
is pointed out.
Several examples of four-dimensional dual pairs are 
constructed where on both sides the geometry of the involved manifolds relies
on del Pezzo surfaces.

\end{abstract}

\bigskip \bigskip
\newpage

\newcommand{\be}{\begin{equation}}
\newcommand{\eq}{\end{equation}}
\newcommand{\ov}{\overline}
\newcommand{\un}{\underline}
\newcommand{\p}{\partial}
\newcommand{\th}{\theta} 
\def\href#1#2{#2}
\def\L{{\cal L}}
\def\M{{\cal M}}
\def\beqa{\begin{eqnarray}}
\def\eeqa{\end{eqnarray}}
\renewcommand{\thesection}{\arabic{section}}
\renewcommand{\theequation}{\thesection.\arabic{equation}}
\parindent1em

\pagenumbering{Roman}
\tableofcontents
\pagenumbering{arabic}
\newpage
\setcounter{page}{1}

%------------------------------------------------------------------------------
\section{Introduction}
%------------------------------------------------------------------------------
The standard model of elementary particle physics has been verified to 
high precision. All fermions in the standard model have been discovered, only 
the Higgs particle is missing, discovering it, one is tempted to say that 
'particle physics is finished'.

However, there are a few conceptual difficulties with the standard model, which
indicate that one should expect new physics beyond. 

For example, one would like to know, why there are nineteen arbitrary 
parameters in the
standard model and how they can be reduced. Also, one should be able to
explain the origin of the three fermion generations and how their masses can
be predicted. 

Another striking question is the gauge hierarchy problem.
It refers to the fact that the weak mass scale, set by the
Higgs mechanism at $m_{weak}=250$ GeV, is not stable under radiativ 
corrections. If gravity is included the hierarchy problem demands for 
explanation of the relativ smallness of the weak scale, compared with the
Planck scale ($m_{Pl}=10^{19}$ GeV). 

Further important questions are the following: why is the standard model 
gauge group $SU(3)\times SU(2)\times U(1)$; why does the cosmological 
constant vanish; why our world is four-dimensional and how gravity is 
included at the quantum level? 

A first step in answering these questions was the introduction of grand
unified theories (GUTs). They are based on gauge symmetry groups
like $SU(5)$ or $SO(10)$. These groups are expected to appear at high 
energies, thus should unify the 
gauge couplings, and expected to be broken at low energies 
to the standard model gauge
group. Further, they explain charge quantization and the relationship 
between quarks and leptons. But, they do not provide answers on the 
hierarchy problem
and other related questions. In particular, they predict the proton 
decay, which has been not observed in nature up to now. 

A next step was the introduction of global supersymmetry (SUSY) which 
essentially solves the hierarchy problem. Due to the existence of 
superpartners for all particles, the quadratic self-mass divergences of 
the scalar Higgs field vanish, and therefore 
the weak scale is stabilized. So far no supersymmetric partners 
have been discovered, so 
supersymmetry, if it exists, has to be broken. Now, the spontaneous 
breaking of global    
supersymmetry leads to the existence of massless fermionic states, the 
Goldstinos, which are also not observed. 

As an alternative one can introduce local supersymmetry, which in contrast
to global SUSY incorporates gravity (SUGRA), avoids the Goldstinos, makes the 
vanishing of the cosmological constant possible, but is like gravity itself
non-renormalizable. 
    
The most promising candidat for solving all these problems is the superstring
theory \cite{GSW},\cite{LBUCH}. A string is an extended object, with a size of 
$\sim 10^{-33}cm$, and replaces the point particles 
as elementary objects.  More precisely, all of the known
elementary particles and gauge bosons can be realized as the different
excitation modes of a string. Among the excitations one finds a spin two
massless excitation, which can be identified as the graviton, 
and moreover, supergravity is incorporated as low energy effective theory.    
Another nice property of string theory is, due to the cancellation of 
anomalies, the dimension of space-time and gauge groups are 
fixed. The cancellation of the conformal anomaly fixes the space-time 
dimension to be 10 for the superstrings and 26 for the bosonic strings.  
The cancellation of gauge and gravitational anomalies fixes the gauge group
to be either $E_8\times E_8$ or $SO(32)$ in ten dimensions. In particular, 
there are only five consistent superstring theories in ten dimensions. The 
type IIA, the type IIB string, the type I with $SO(32)$ gauge group and two 
heterotic string theories with $E_8\times E_8$ resp. $SO(32)$ gauge group.
The common features of all five string theories are: they are space-time
supersymmetric and contain a graviton, an antisymmetric tensor field and the
so called dilaton in their massless spectrum. Moreover, the vacuum expectation
value of the dilaton controlls (at the tree level) the gauge couplings. 
More precisely, the bosonic spectrum of the type I and type II theories appears
in two sectors, the Ramond-Ramond (RR) or Neveu-Schwarz-Neveu-Schwarz (NS-NS)
sector, depending on the boundary conditions of the worldsheet fermions, which 
are periodic (RR) resp. anti-periodic (NS-NS). In particular, the type IIA and
IIB string theories contain only closed strings and have two gravitinos, with
opposite chiarlity in IIA and the same chirality IIB, whereas both lead to
$N=2$ SUGRA in their low energy limit. In contrast, type I contains, in 
addition to closed also open strings, has one gravitino and leads to $N=1$
SUGRA in the field theory limit. The two heterotic string theories contain
closed strings, they have one gravitino and lead also to $N=1$ SUGRA in the
field theory limit. 

Altough strings are the most promising candidates for grand unified model 
building, there are a few inherent stringy problems. First of all, since the
critical dimension for consistent string theories is ten, one has to compactify
six of them. This means that a vacuum configuration must be specified, in 
order to make contact with the four-dimensional world.
Second, if string theory claims to be a unique 
fundamental theory, why there are five consistent theories in ten dimensions?
Third, string theory incorporates not only the right elementary excitations   
and appropriate gauge groups for successful GUT predictions, they also
contain gravity and thus should make predictions for the gravitational 
coupling strength, i.e. for Newton's constant, predictions which fail in
perturbative heterotic string theory. 
Not only the latter fact, but also a desirable 
understanding of supersymmetry breaking leads to a fourth point, a 
nonperturbative formulation of string theory, including nonperturbative 
effects at strong coupling. The necessity of a nonperturbative formulation
is also supported by the fact, that the specification of the vacuum 
configuration can be understood dynamically. 

Something one can do, at least, is to combine some demands as: maintaining 
conformal invariance, unbroken supersymmetry and vanishing of the
gauge anomanly (if ten-dimensional gauge groups are involved) 
in order to restrict possible vacuum configurations. 
It has been shown that this favors vacuum configurations based on 
Calabi-Yau compactifications, so that one ends up with vacua which
are products of the four-dimensional Minkowski space $M_4$ and a complex
three-dimensional, Ricci-flat K\"ahler manifold $X$, i.e. $X$ has vanishing
first Chern-class, and so preserving conformal invariance.           
Imposing the condition of Ricci-flatness to K\"ahler manifolds of complex
dimension $n$, reduces the holonomy group $U(n)$ to $SU(n)$, which is stated
by a theorem of Calabi and Yau. 

In fact, compactifications, say, of the 
heterotic string on a Calabi-Yau threefolds leads to vacua with $N=1$
supersymmetry in four dimensions. The gauge anomaly is cancelled by embedding
the holonomy group in the gauge group $E_8\times E_8$, which is 
called 'the standard embedding' (i.e. the $SU(3)$ spin connection of $X$ is
identified with an $SU(3)$ subgroup of $E_8$), thus leaving an 'hidden' $E_8$
and an $E_6$ as viable GUT group.

Now, why are there five consistent string theories in ten dimensions? 
This question is expected to be answered due to the establishment of various 
types of duality symmetries between different string vacua, which were
inspired by duality considerations in supersymmetric field theories. 
More precisely, Montonen and Olive stated that the $N=4$ supersymmetric 
Yang-Mills theory admits a weak-strong coupling duality, exchanging 
elementary (electric charged) with solitonic (magnetic charged) states.
In string theory, a similar duality was observed in heterotic string
compactifications on a six-dimensional torus, which leads to $N=4$ 
supersymmetry in four dimensions. This duality, called $S$-duality 
\cite{FILQ}, is also
a strong-weak coupling duality and so a nonperturbative symmetry of 
this theory. Also, the uncompactified type IIB superstring possesses such a
symmetry, transforming states (BPS p-branes) from the Neveu-Schwarz sector
into states (Dirichlet p-branes) of the Ramond sector.      

All these dualities are only symmetries of one string theory, however, it
has been shown that there are also duality symmetries relating the weak 
coupling region of one string theory to the strong coupling region of another
and vice versa. The duality is called string-string duality. For example, the
type I theory is conjectured to be dual to the $SO(32)$ theory in 
ten dimensions \cite{W95}, \cite{PoW}. Another example is the heterotic string
compactified on $T^4$ which is conjectured to be dual to the type IIA string on
$K3$ in six dimensions \cite{HT}, \cite{HS}. Kachu and Vafa \cite{KachuV} 
stated the duality between the heterotic string compactified on $K3\times T^2$ 
and the type IIA string on Calabi-Yau threefold which leads to $N=2$ 
supersymmetry in four dimensions. 

Tests of these dualities would certainly involve computations at strong
coupling, but since string theory is only formulated perturbatively, this
seems to be offhand impossible. Fortunately supersymmetry leads to 
certain non-renormalization theorems and therefore some weak coupling 
calculations are exact, they do not require nonperturbative corrections. 
So one has to look for certain quantities which are uncorrected. In theories 
with extended supersymmetry, such quantities are related to BPS states. 
BPS states saturate the Bogomolny bound between mass and charge; their mass
is completely determined in terms of their charge. For example, they are
relevant in tests of the nonperturbative $SL(2,{\bf Z})$ duality of the 
$N=4$ theory in four dimensions, obtained by compactification of the 
heterotic string on $T^6$ \cite{SEN}, and in the study of one-loop 
threshold corrections to gauge and gravitational couplings in $N=2$ 
heterotic string compactifications \cite{HM}. 

In particular, tests could 
successfully performed in a broad class of type II compactifications on 
Calabi-Yau threefolds and heterotic vacua on $K3\times T^2$ with $N=2$
supersymmetry in four dimensions (for a review c.f. \cite{LURE}). 
Further it has been shown that
using these $N=2$ string duality symmetries, results
about nonperturbative effects in string theory and field theory
can be derived, for example, the computation of the nonperturbative
heterotic $N=2$ prepotential or nonperturbative effects in $N=2$
field theory like Seiberg/Witten \cite{SW}, \cite{KKLMV}, \cite{KLMVW}.   

It has also been shown that solitonic objects, which are given in string
theory by membranes and p-branes, have to be included in order to successfully
test string-string duality \cite{HT}, \cite{W95}. 

Due to the duality describtions 
between different string theories, it is now believed that all five string
theories are perturbative expansions of one fundamental underlying theory,
called M-theory. It is not clear up to now what M-theory is, but at low 
energies it should reduce to eleven-dimensional supergravity.   
The supporting arguments which led to the existence of M-theory are twofold.

First, the type IIA string has  
$N=2$ supergravity as ten-dimensional field theory limit. It has been 
shown \cite{Camp}
that $N=2$ supergravity in ten dimensions can be obtained by compactification
of eleven-dimensional supergravity on $S^1$ where the radius of $S^1$ is 
related to the type IIA string coupling as $R_{11}\sim g_{IIA}^{\frac{2}{3}}$.
Witten has argued \cite{W95} that in the strong coupling limit  
(i.e. $R_{11}\rightarrow \infty$) the type IIA string theory approaches an 
eleven-dimensional (Lorentz invariant) theory whose low energy limit 
is eleven-dimensional supergravity. 

Second, it has been shown \cite{W142},\cite{W70} that the strong 
coupling limit of the 
heterotic string can be described by an eleven-dimensional theory 
compactified on $S^1/{\bf Z}_2$ where the $E_8\times E_8$ gauge fields live
on the two ten-dimensional boundaries respectively.     
As a consequence of the strong coupling scenarion, the predictions
for Newton's constant are relatively close to the actual value.  

As a next step one can try to derive the type IIB and SO(32) resp. type I 
string theory from M-theory and therefore try to explain the geometric 
origin of the $SL(2,{\bf Z})$ symmetry of type IIB theory. Specifically, one
can argue that type IIB compactified on $S^1$ has $SL(2,{\bf Z})$ symmetry
in nine dimensions and is T-dual to type IIA theory compactified on $S^1$
thus M-theory on $T^2$ is dual to type IIB on $S^1$ and the $SL(2,{\bf Z})$
symmetry of type IIB becomes the U-duality for type IIA on $S^1$ \cite{S143}. 
The $SL(2,{\bf Z})$ symmetry is just the group of global diffeomorphisms of 
$T^2$ \cite{A154}, \cite{S086}. But in going to ten dimensions, one has to 
take the zero size limit of $T^2$ and therefore looses the geometric 
description of M-theory (similarly, one can show for SO(32) resp. type I 
theory). 

Now, taking $SL(2, {\bf Z})$ symmetry in ten dimensions leads Vafa \cite{V}
to the 
introduction of a twelve-dimensional theory, called F-theory, which 
compactified on $T^2$ leads to type IIB in ten dimensions (or compactified 
on $T^2/{\bf Z}_2$ leads to SO(32) theory whose strong coupling limit 
is described by a weakly coupled ten-dimensional type I theory 
\cite{W95}, \cite{PoW}). In particular, the $SL(2,{\bf Z})$ symmetry of the 
dilaton-axion field $\tau$ (a complex field constructed with the R-R axion 
$\tilde\phi$ and 
the NS-NS dilaton $\phi$) can be interpreted as the modular invariance 
of $T^2$ and so the $SL(2,{\bf Z})$ symmetry of type IIB theory is
geometrized. 

In standard perturbative compactifications of type IIB string theory the 
dilaton-axion field $\tau$ is constant on the internal manifold $B$. 
As it has been shown \cite{V}, \cite{MV1}, \cite{MV2}, compactifications of 
F-theory to lower dimensions can be 
formulated as type IIB theory on $B$ with varying $\tau$ over $B$. 
Regarding F-theory in this way, avoids the difficulties in formulating a 
consistent twelve-dimensional theory which would have a metric of (10,2) 
signature. 
The fact that $\tau$ is identified with the complex structure modulus of 
$T^2$ and varies over $B$, makes one consider elliptically fibered manifolds,
thus preserving the $SL(2,{\bf Z})$ symmetry. In a generic F-theory
compactification the elliptic fiber can degenerate over a codimension one 
subspace $\Delta$ in the base and therefore non-trivial closed cycles in $B$ 
can induce no-trivial $SL(2,{\bf Z})$ transformation on the fiber. Thus the
dilaton is not constant, it can jump by an $SL(2,{\bf Z})$ transformation, a
fact which has no explanation in perturbative string theory. Moreover, it
has been shown that the non-trivial monodromy around closed cycles signals
the presence of magnetically charged D-sevenbranes at $\Delta$, filling 
the uncompactified space-time \cite{V}.     

So, besides the explanation of the $SL(2,{\bf Z})$ symmetry of the 
ten-dimensional type IIB theory, F-theory 
provides new nonperturbative type IIB
vacua on D-manifolds with varying coupling over the internal space.     

Moreover, F-theory also provides an alternative description of heterotic 
string compactification in purely geometrical terms. In particular, it has 
been shown \cite{V}, \cite{MV1}, \cite{MV2} that F-theory
compactified to eight dimensions on $K3$ is dual to the heterotic string
on $T^2$. Once the eight-dimensional duality is established, one can extend
adiabatically \cite{VW} over a base manifold $B$ and thus obtain lower
dimensional dualities. This has been done for one-dimensional $B$ leading 
to $N=1$ heterotic/F-theory dual pairs in six dimensions.
Of even greater phenomenological interest is the investigation
of string duality symmetries in four-dimensional string vacua with
$N=1$ space-time supersymmetry. Therefore one compactifies the heterotic 
string on an elliptically fibered Calabi-Yau threefold 
with a vector bundle, which
breaks part of the $E_8\times E_8$ group to some possibly viable GUT group.
This is conjectured to be dual to F-theory on elliptically fibered 
Calabi-Yau fourfold. The unbroken heterotic gauge group is expected to
be localized on a locus of degenerated elliptic fibers in the fourfold. 
Furthermore, it has been shown that consistent F-theory compactifications
on Calabi-Yau fourfold require a number of space-time filling threebranes 
which are localized at points in the base $B$ of the elliptic fourfold 
(thus, their worldvolume is a four-dimensional submanifold of space-time and 
given by ${\bf R}^4\times p_i$, where $p_i$ are the points in $B$). The number
of such threebranes was determined in \cite{SVW} by observing that the SUGRA 
equations have a solution only for a precise number of such threebranes 
(more correctly, this number has been shown to be proportional to the 
Euler characteristic of the fourfold). Under duality these threebranes 
should turn into heterotic fivebranes \cite{FMW}.
 
Now, if the map between the heterotic string description and F-theory 
can be made precise, important nonperturbative information, like 
the computation of nonperturbative $N=1$ superpotentials \cite{EW}, 
\cite{DGW}, \cite{Gan}, \cite{CL} 
and supersymmetry breaking, or the
reformulation of many effects in $N=1$ field theory \cite{KKV},
\cite{KV}, \cite{KV90}, \cite{OV80} are expected.
In addition, the way in which transitions among $N=1$ heterotic vacua,
possibly with a different number of chiral multiplets, take place, 
is a very important question.
\\

Our primary topic is the study of $N=1$ heterotic/F-theory
duality in four dimensions. In the first part we will review
and develop some methods and tools needed for studing four-dimensional
heterotic/F-theory duality.

Let us be more precise!

In section 2, we review the constraints for consistent $E_8\times E_8$
heterotic string compactification and give an example of how the massless
spectrum is obtained. 

In section 3, we first specify the class of Calabi-Yau manifolds 
which we will consider in the remaining part. 

In addition to the specification of the manifold, we have to specify a stable
vector bundle, in order to compactify the heterotic string, this breaks 
part of the $E_8\times E_8$ gauge symmetry. 

Actually, there are three methods for constructing stable 
vector bundles over
elliptic fibrations: the parabolic, the spectral cover and the construction
via del Pezzo surfaces which are explained and developed in \cite{FMW} and 
\cite{FMW2}.

We adopt here the parabolic construction, however, we compare 
our results to those obtained from the spectral cover construction. 

We consider a $G=SU(n)$ vector bundle, with $n$ odd, which 
determines a rank $n$ complex vector bundle $V$ of trivial determinant. 

We first review, for our purposes, some facts of the 
parabolic construction for $V$ an $SU(n)$ vector bundle, then we will recall
what is known for $n$-even which was discussed in \cite{FMW} and we
will extend this to the case if $n$ is odd. In contrast 
to \cite{FMW} we do not 
focus on a $\tau$-invariant point in the moduli space of $SU(n)$ bundles. 
This allows us to determine the net number $N_{gen}$ of generations of 
chiral fermions, i.e.
($\sharp$ generations $-$ $\sharp$ antigenerations)
in the observable sector of the 4D unbroken gauge group. In contrast to
the bundles at the $\tau$-invariant point, which have $n$ even and 
a modulo 2 condition for $\eta$, our bundles have $n$ odd and 
$\pi_*(c_2(V))=\eta$ divisible by $n$.
In particular it will be shown that for a certain choice of the twisting 
line bundle on the spectral cover both approaches agree. 

In the last part of section 3, we will determine the number of bundle moduli
for $SU(n)$ and $E_8$ vector bundles on elliptically fibered threefolds using
technics (a character-valued index) computed and applied in \cite{FMW} for
$SU(n)$ vector bundles.       

In section 4, we review the framework of F-theory, the constraints for 
consistent four-dimensional F-theory compactification on Calabi-Yau fourfold
and recall the adiabatic principle, needed to establish the duality between
the heterotic string and F-theory.  

In section 5.1 and 5.2, we review some facts about eight-and six-dimensional
het/F-theory duality, some examples for the six-dimensional case are presented.
In particular, section 5.2 provides us with the
necessary tools we will later use to construct four-dimensional 
heterotic/F-theory models from six-dimensional ones.

In section 5.3, we show  
that in dual $N=1$ heterotic/F-theory vacua the number of heterotic fivebranes
necessary for anomaly cancellation matches the number of F-theory
threebranes necessary for tadpole cancellation. This extends to the general 
case the work of Friedman, Morgan and Witten, who treated the case of 
embedding a heterotic $E_8\times E_8$ bundle, leaving no unbroken gauge group,
where one has a smooth Weierstrass model on the F-theory side.
 
Further, we compare the spectrum of chiral multiplets obtained in 
four-dimensional heterotic string compactification with vector bundle $V$
with the spectrum of chiral multiplets obtained from F-theory compactification
on Calabi-Yau fourfold. We find a complete matching of both spectra. 

In section 6.1, we consider F-theory on smooth fourfolds $X^4_k$ elliptic
over $dP_k \times {\bf P}^1$ as an example.

In section 6.2, we study a class of Calabi-Yau threefolds elliptic
over $dP_k$ and use the index formula for the number of moduli of
a general $E_8 \times E_8$ bundle and compare the spectra with those
obtained from the fourfolds $X^4_k$.

In section 6.3, we consider the ${\bf Z}_2$ modding of the $N=2$ models
described by F-theory on $T^2 \times X^3_n$, where $X^3_n$ are the well
known threefold Calabi-Yau over $F_n$.

In section 6.4, we make the corresponding modding in the dual heterotic $N=2$
model on $T^2 \times K3$ with instanton embedding $(12+n,12-n)$.

In section 6.5, we consider the standard embedding, which corresponds to
a purely perturbative compactification (without fivebranes). We show that
the number of threebranes also vanishes on the F-theory side, and study
the 6D analog of the complete Higgsing process. We find evidence that the 
4D spectrum can be obtained from compactification of the heterotic
string on $K3\times T^2/(\sigma, -1)$.   

\newpage

%------------------------------------------------------------------------------
\section{$N=1$ Heterotic Vacua}
%------------------------------------------------------------------------------
\subsection{Perturbative Vacua} 
%------------------------------------------------------------------------------
If we study the $E_8\times E_8$ heterotic string in a flat Minkowskian 
background, then the cancellation of the conformal anomaly forces us to 
choose a ten-dimensional spacetime. Requiring unbroken $N=1$ spacetime 
supersymmetry of the effective theory, which arose from compactification
on $M_4\times Z$, is equivalent to requiring that 

\be  
<\Omega|\delta\Psi|\Omega>=0
\eq
for all fundamental fermionic fields in the spacetime Lagranian where 
$|\Omega>$ is the vacuum state of the supersymmetric field theory and $\delta
\Psi$ denotes the variation of $\Psi$ under a supersymmetry transfromation.
Since the effective theory that arises from heterotic string theory is 
$N=1$ supergravity coupled to super Yang-Mills theory $\Psi$ can be a 
gravitino $\psi_i$, a gluino $\chi ^a$ or a dialtino $\lambda$. The gravitino
transforms under supersymmetry transformation as 

\be
\delta\psi_i = \frac{1}{\kappa}D_i\eta+\frac{\kappa}{32g^2\phi}(
                \Gamma^{jkl}_i-9\delta^j_i\Gamma^{kl})\eta H_{jkl} 
\eq
where $\Gamma^{jkl}_i$ are anti-symmetrized products of 
gamma matrices, $\eta$ denotes the Grassman parameter, $\phi$ is the dilaton, 
$H$ is the field strength for the antisymmetric tensor field $B$ and 
$g$, $\kappa$ are the gravitational and gauge coupling constants. 
$g$ and $\kappa$ are not independent parameters, they can be reabsorbed in 
the dilaton so that the only independent parameter in the theory is the VEV 
of $\phi$.  

With the ansatz $H=d\phi=0$, $\delta\psi_i=0$ tells us that the manifold $K$
must admit a covariant constant spinor field $\eta$, i.e. $K$ must be a
K\"ahler manifold with vanishing first Chern class and a metric of $SU(3)$
holonomy, i.e. $K$ is a Calabi-Yau manifold.\\
The supersymmetry transformation for the gluinos $\chi^a$ are given by

\be
\delta\chi^a = -\frac{1}{4g\sqrt{\phi}}\gamma^{ij}F^a_{ij}\eta    
\eq
and $\delta\chi^a=0$ leads to two conditions: 

\be
F_{ij}=F_{\bar{i} \bar{j}}=0 
\eq
tells us that the vacuum gauge field $A$ is a holomorphic connection
on a holomorphic vector bundle $V\rightarrow K$ where $V$ must be some 
subbundle of the $E_8\times E_8$ bundle. The second condition   

\be
g^{i\bar{j}}F_{i\bar{j}}=0
\eq
is the so called Donaldson-Uhlenbeck-Yau equation for the connection $A$ 
which has a unique solution precisely if the integrability condition 
$\int_K \Omega^{n-1}\wedge c_1(V)=0$ is satisfied and $V$ is a  
stable \footnote{Let $Z$ be a compact K\"ahler manifold of complex 
dimension $d$ with 
K\"ahler form $\Omega$ and let $V$ be a holomorphic rank $n$ vector bundle
on $Z$. One defines the normalized degree (or slope) of $V$ with respect
to $\Omega$ as the real number
$\mu(V)=\frac{1}{rank(V)}\int_Z c_1(V)\wedge \Omega^{d-1}$.
A holomorphic rank two vector bundle over $Z$ is {\it stable} with respect to
$\Omega$ ($\Omega$-{\it stable}) if, for any line bundle 
$F$ on $Z$ there exists a nonzero map $F\rightarrow V$, one has
$\mu(F)< \mu(V)$
and $V$ is called {\it semistable} if $\mu(F)\le\mu(V)$ holds. 
For arbitrary rank ${\it stability}$ resp. ${\it semistability}$ are 
similarly defined; one replaces $F$ by a coherent analytic subsheaf ${\cal F}$
of the sheaf of holomorphic sections of $V$ such that 
$0< rank {\cal F}< rank V$. However, we are interested in vector bundles 
over Calabi-Yau manifolds for which $\mu(V)=0$. We need $H^0(Z,V)=H^3(Z,V)=0$,
which means that $V$ has no sections. Otherwise, a nonzero 
element of $H^0(Z,V)$ 
defines a mapping from ${\cal O}\rightarrow V$ from the trivial line bundle
into $V$ but since the trivial line bundle has rank one, this would violate 
the inequality.} bundle or a dircet sum of 
stable bundles. This was shown by Donaldson
\cite{D} for the case of $K$ being complex two-dimensional and by 
Uhlenbeck and Yau \cite{UY} for higher dimensional cases. 
Since the K\"ahler 
form $\Omega^n$ of $K$, which is the volume form of $K$, has nonzero integral 
over $K$, we are restricted to cases where the first Chern class $c_1(V)$ 
vanishes identically which is a stronger condition than $c_1(V)=0$ (mod 2),
the obstruction to guarantee that the bundle $V$ admits spinors.

Finally, the dilatino transforms as

\be
\delta\lambda = -\frac{1}{\sqrt{2}\phi}(\Gamma\cdot{\p}\phi)\eta
                 +\frac{\kappa}{8\sqrt{2}g^2\phi}\Gamma^{ijk}\eta H_{ijk}
\eq 
and with the above ansatz of vanishing three form $H$ and constant dilaton 
$\phi$, $\delta\lambda=0$ is satisfied. 
Furthermore, we have to consider the Bianchi identity for $F$ and $H$ which
is essentially a topological condition and given by

\be
dH=trR\wedge R- trF\wedge F
\eq
and simplifies in our case to $trR\wedge R=trF\wedge F$ or
equivalently to 

\be
c_2(TK)=c_2(V)
\eq
where $TK$ denotes the tangential bundle of $K$.

In summary, we have found that in order to compactify the heterotic string,
one has to specify a Calabi-Yau manifold $Z$ and a stable, holomorphic 
vector bundle $V\rightarrow Z$ satisfying the conditions

\be
c_1(V)=0, \ \ \ c_2(V)=c_2(TZ).
\eq

Let $V$ be a $SU(n)$ vector bundle over $Z$.  
The unbroken space-time gauge group will be the maximal subgroup of 
$E_8\times E_8$
that commutes with the vacuum gauge field. For example, a $SU(n)$ vector bundle
embedded in one $E_8$, with $n=3,4$ or $5$, leads to an unbroken $E_6$, 
$SO(10)$ resp. $SU(5)$ gauge group (times a "hidden" $E_8$ which couples
only gravitationally).  

The spectrum of charged matter is directly related
to properties of $Z$ and $V$. So, let us obtain the spectrum of 
massless fermions! We start in ten dimensions with the Dirac equation
\beqa
iD_{10}\Psi=0=i(D_4+D_Z)\Psi
\eeqa
further making the ansatz 
\beqa
\Psi=\psi(x)\phi(y)
\eeqa
where $x$ and $y$ are coordinates on $Z$ respectively $M_4$. If $\psi$ is an 
eigenspinor of eigenvalue $m$ then it follows
\beqa
iD_Z\psi=m\psi
\eeqa
and so we get
\beqa
(iD_4+m)\phi=0.
\eeqa
So we learn that $\psi$ looks like a fermion of mass $m$, to a four-dimensional
observer. Thus, massless four-dimensional fermions are in one to one
correspondence with zero modes of the Dirac operator on $Z$. The charged 
four-dimensional fermions are obtained from ten-dimensional ones,
which transform under the adjoint of $E_8$. 

Now since massless fermions in four dimensions are related to the zero 
modes of the Dirac operator on $Z$, they can be related to
the cohomology groups $H^k(Z,V)$. Now, the index of the 
Dirac operator can be written as 
\beqa
\chi(V)=index(D_V)=\int_Z td(Z)ch(V)=\frac{1}{2}\int_Z c_3(V).
\eeqa 
On the other hand we have 
\beqa
\chi(V)=\sum_{i=0}^{3}(-1)^k h^k(Z,V)
\eeqa
and if we recall that $h^0(Z,V)=h^3(Z,V)=0$ for stable bundles, we can write 
\beqa
h^2(Z,V)-h^1(Z,V)=\frac{1}{2}\int_Z c_3(V).
\eeqa  
and by Serre duality on $Z$: $h^2(Z,V)$ is dual to $h^1(Z,V^*)$. 
The index measures the net number of generations $N_{gen}$ of chiral fermions,
thus we get
\beqa
N_{gen}=\frac{1}{2}c_3(V)
\eeqa  
In addition one has to take into account a number of bundle moduli 
$h^1(Z,End(V))$, which are gauge singlets.  

%------------------------------------------------------------------------------
\subsection{Non-perturbative correction}
%------------------------------------------------------------------------------
So far we have reviewed the conditions for perturbative heterotic string
compactifications. Now, Duff, Minasian and Witten have argued, that the
perturbative anomaly cancellation condition can be modified in the presence 
of nonperturbative fivebranes in compactifications to six dimensions 
\cite{DMW} 
\beqa
n_1+n_2+n_5=c_2(Z).
\eeqa
Further, Friedman, Morgan and Witten \cite{FMW} have shown
that a similar 
relation appears in 
four-dimensional heterotic string compactifications on elliptically
fibered Calabi-Yau threefolds. 
More precisely, the consistency of F-theory compactification on Calabi-Yau
fourfolds requires the presence of a number $n_3$ of threebranes in the 
vacuum \cite{SVW}, which should turn into fivebranes under duality with
the heterotic string \cite {FMW} which wrap over fibers of 
$\pi: Z\rightarrow B_2$. The question was: why does a precise number of 
fivebranes appear in the heterotic vacuum? They showed that this is 
explained by anomaly cancellation. They considered an 
$E_8\times E_8$ vector bundle $V=V_1\times V_2$ which is determined by its
fundamental characteristic class $\lambda(V_1)$ and $\lambda(V_2)$
\footnote{$\lambda(V)=\frac{1}{8\pi^2}tr F\wedge F$, taking the trace in the 
adjoint representation leads to $\lambda(V)=c_2(V)/60$} with 
\beqa
\lambda(V_i)=\sigma\pi^*(\eta_i)+\pi^*(\omega)
\eeqa
where $\eta_i\in H^2(B_2,{\bf Z})$ and $\omega\in H^4(B_2,{\bf Z})$, 
in particular, $\eta_i$ is arbitrary but $\omega$ is determined in terms 
of $\eta_i$. Then, it was shown that the push-forward of 
$\lambda(V_i)=c_2(TZ)$ leads to ($TZ$ denotes the tangent bundle to $Z$) 
\beqa
\pi_*(c_2(TZ))-\pi_*(\lambda(V_1))-\pi_*(\lambda(V_2))=h[p].
\eeqa
with $h\in{\bf Z}$ and $[p]$ is the class of a point 
and therefore the pullback of 
$\lambda(V_i)=c_2(TZ)$ is in error by a cohomology class of $B_2$. 
This leads to the general anomaly cancellation condition with fivebranes  
\beqa
\lambda(V_1)+\lambda(V_2)+[W]=c_2(TZ)
\eeqa
where $[W]=h[F]$ denotes the cohomology class of the fivebranes and
$[F]=\pi^*([p])$ is the class dual to the fibre. 

Furthermore, it was shown \cite{FMW} that a choice of $\eta_i$, leads 
to a prediction for $h$, which agrees with the number of F-theory threebranes 
$n_3$ and so $h=n_5$. 
\\
{\bf{Note:}} We will presently show (c.f. section 5) 
that $n_3=n_5$ holds 
also for general $SU(n)$ vector bundles!

\newpage
%------------------------------------------------------------------------------
\section{Calabi-Yau spaces and Vector bundles}
\resetcounter
%------------------------------------------------------------------------------
In section 3.1, we state a few facts about a class of 
elliptically fibered Calabi-Yau
manifolds, which can be represented by a smooth Weierstrass model. These
manifolds are used in the heterotic/F-theory duality.\\ 
In section 3.2, we  
construct a certain class of $SU(n)$ vector bundles, with $n$ odd, and
compute their third Chern-class $c_3(V)$, which is related to the net amount 
of chiral matter.\\     
In section 3.3, we derive the number of bundle moduli for the
$E_8$ bundle, for the tangent bundle of a smooth elliptic Calabi-Yau threefold,
and review the computation for $SU(n)$ bundles, which has been performed 
in \cite{FMW}.
%------------------------------------------------------------------------------

\resetcounter
\subsection{Elliptic fibrations}
%------------------------------------------------------------------------------
Let $Z$ be an elliptic fibered Calabi Yau 
$\pi: Z\rightarrow B$ and let $\sigma$ be a section of $\pi$ such that
$\pi\circ\sigma=id_B$. 

In the following we are interested in the cases that $B$ being complex 
$1,2$ or $3$ dimensional denoted by $B_1, B_2$ and $B_3$ respectively 
which leads then to a elliptic fibered Calabi-Yau twofold (the $K3$), 
a Calabi-Yau three- respectively fourfold.

$Z$ can be described by a Weierstrass equation 

\be
y^2z = x^3+g_2xz^2+g_3z^3
\eq
which embeds $Z$ in a ${\bf P}^2$ bundle $W \rightarrow B$; $g_2$ and $g_3$
are polynomilas on the base. 
The ${\bf P}^2$ bundle is the projectivization of a vector bundle  
${\bf P}(\L^2 \oplus \L^3 \oplus {\cal O}_B)$ with $\L$ being a line bundle 
over $B$. Since the canonical 
bundle of $Z$ has to be trivial we get from $K_Z=\pi^*(K_B+\L)$ the condition 
$\L=-K_B$. Further we can think of $x,y$ and $z$ as homogeneous coordinates 
on the ${\bf P}^2$ fibers, i.e. they are sections of ${\cal O}(1)\otimes 
K_B^{-2}$, ${\cal O}(1)\otimes K_B^{-3}$ and ${\cal O}(1)$. Furthermore 
$g_2$ and $g_3$ are sections of $H^0(B_3, K_B^{-4})$ and $H^0(B_3, K_B^{-6})$ 
respectively \cite{FM}. Thus $Z$ as defined through the Weierstrass 
equation is a section of ${\cal O}(1)^3\otimes K_B^{-6}$.
The section $\sigma$ can be thought of as the point 
at infinety in ${\bf P}(K_B^{-2} \oplus K_B^{-3} \oplus {\cal O}_B)$ defined 
by $x=z=0, y=1$. 

Also we have to note that such an elliptic fibration has a ${\bf Z}_2$
symmetry which acts as multiplication by -1 on the canonical bundle
\footnote{The canonical bundle of the elliptic fibers is generated by the 
differential form $dx/y$.} of the elliptic fibers of $\pi$ while acting
trivially on the canonical bundle of the base. If $Z$ is represented as
above, the ${\bf Z}_2$ symmetry is generated by the transformation
\beqa
\tau:y\rightarrow -y. 
\eeqa
The descriminant $\Delta$ of $\pi$ is given by 
\be
\Delta(g_2, g_3)=g_2^3-27g_3^2
\eq
which is a section of $K_B^{-12}$. The $j$-invariant of $Z$  
\be
j_Z=1728\frac{g_2^3}{\Delta}
\eq
is a quotient of two sections of $K_B^{-12}$ and thus a meromorphic
function on $B$. For any point $p$ on $B$ with $\pi^{-1}(p)=E_p$ nonsingular
fiber $j_Z(p)=j(E_p)$ is the usual $j$-invariant.

If $\Delta=0$ then the elliptic fiber is singular.
The type of singular fibre is determined by the order of vanishing of $g_2$,
$g_3$ and $\Delta$ at codimension one in the base. 
Let us denote by $F, G$ and $D$ the classes of the divisors
associated to the vanishing of $g_2, g_3$ and $\Delta$ respectively, since 
$X$ being Calabi-Yau we have $F = -4K_B$, $G = -6K_B$ and $D = 
-12K_B$. 
{\subsubsection{Characteristic classes}}
As noted above $x$, $y$ and $z$ can be thought of as homogeneous coordinates
on the ${\bf P}^2$ bundle $W$ over $B$ respectively as sections of line bundles
whose first Chern classes we denote by $c_1({\cal O}(1)\otimes K_B^{-2})=
r+2c_1$,  $c_1({\cal O}(1)\otimes K_B^{-3})=r+3c_1$ and $c_1({\cal O}(1))=r$.
The cohomology ring of $W$ is generated over the cohomology ring of $B$ by the 
element $r$ with the relation $r(r+2c_1(\L))(r+3c_1(\L))=0$ which expresses 
the fact that $x$, $y$ and $z$ have no common zeros. 

Since $Z$ is defined by the vanishing of a section of ${\cal O}(1)^3\otimes 
K_B^{-3}$, which is a line bundle over $W$ with first Chern class $3r+6c_1$,
any cohomology class on $Z$ that can be extended to one over $W$, can be 
integrated over $Z$ by multiplying it by $3r+6c_1$ and then integrating 
over $W$, i.e. multiplication by $3(r+2c_1)$ can be understood as restricting 
$W$ to $Z$.

The relation $r(r+2c_1(\L))(r+3c_1(\L))$ can then be simplified to 
$r(r+3c_1)=0$ in the cohomology ring of $Z$. Now the total Chern class of the 
tangent bundle of $W$ is 
\be
c(W)=c(B)(1+r)(1+r+2c_1(\L))(1+r+3c_1(\L))
\eq
where $c(B)$ denotes the total Chern class of the tangent bundle of $B$. 
The total Chern class of $Z$ is then obtained by dividing $c(W)$ by 
$(1+3r+6c_1(\L))$
\be
c(Z)=c(B)\frac{(1+r)(1+r+2c_1(\L))(1+r+3c_1(\L))}{1+3r+6c_1(\L)}
\eq
Expanding $c(Z)$ leads to the Chern classes of $Z$. To obey the Calabi-Yau 
condition $c_1(\L)=c_1(B)=c_1$. For the cases of $Z$ being complex
$2,3$ or $4$ dimensional (denoted by $X_2$, $X_3$, $X_4$) resp. with bases 
$B_1$, $B_2$ and $B_3$ we find (using $r^2=-3rc_1$)
\beqa
B_1:\ \ c_2(X_2)&=&4rc_1 \\
B_2:\ \ c_2(X_3)&=&c_2+11c_1^2+4rc_1 \\
        c_3(X_3)&=&-20rc_1^2 \\
B_3:\ \ c_2(X_4)&=&c_2+11c_1^2+4rc_1 \\
        c_3(X_4)&=&c_3-20rc_1^2-c_1c_2-60c_1^3 \\
        c_4(X_4)&=&4rc_1c_2+120rc_1^3 .
\eeqa
The Euler characteristic of $X_i$ is given by
\be
\chi(X_i)=\int_{X_i}c_i.
\eq
$z$ is a section of the line bundle ${\cal O}(1)$ which is defined
on the total space of $W$. The divisor $z=0$ which is dual to the 
cohomology class $r$ intersects the generic fiber $E_p$ in three 
points, i.e. $r\cdot E_p=3$. Hence, first intergrating over the fibers
and then over $B$ leads to the Euler characteristics for $X_2$, $X_3$ and
$X_4$
\beqa
\chi(X_2)&=&12\int_{B_1}c_1 \\
\chi(X_3)&=&-60\int_{B_2}c_1^2 \\ 
\chi(X_4)&=&12\int_{B_3}c_1(c_2+30c_1^2). 
\eeqa\\  
{\bf{Comment:}} Another way to derive $\chi(X_3)$ is as follows: Recall, that
the discriminant $\Delta$ describes a codimension one sublocus in the base $B$.
In case that $B$ is two-dimensional, $D$ is a curve in $B_2$. The class of 
the curve is given by $D= -12K_{B_2}$. Furthermore, we have the classes
$G=-4K_{B_2}$ and $F=-6K_{B_2}$. Now, we are interested in $\chi(X_3)$ which
is described by a smooth Weierstrass model, that tells us that we can only 
expect type I (and II) singular fibers over $D$ which contribute to
$\chi(X_3)$. Now, the singularity type of the singular fiber tells us the 
deviation of the Euler characteristic from the generic elliptic fiber, which 
has $\chi(T^2)=0$. So, we should expect that 
$\chi(X_3)=\chi(sing.fiber)\chi(D)$. Since $D$ is a curve we
have $-D(D+K_{B_2})=-132c_1^2$ (where $c_i=c_i(B_2)$). But $D$ 
itself can be singular! This happens at those points where the divisors 
associated to the classes $F$ and $G$ collide, and this happens at
$F\cdot G=24c_1^2$ points. At these points $D$ develops a cusp and the 
elliptic fibre will be of type II. Using the standard Pl\"ucker formula,
which takes the double points into account (c.f. \cite{GHarris}), we get
$\chi(\tilde{D})=-132c_1^2+2(24c_1^2)$, and for $\chi(X_3)$ we get
\beqa
\chi(X_3)&=&1(-84c_1^2-24c_1^2)+2(24c_1^2)\nonumber\\
&=&-60c_1^2.
\eeqa\\ 
{\bf {Remarks on the base:}} At this point we review some 
facts about $B_3$, which we will need for 
establishing heterotic/F-theory duality following \cite{FMW}.
Let $\sigma:B_3\rightarrow B_2$ be a ${\bf P}^1$ bundle which is the 
projectivization ${\bf P}(Y)$ of a vector bundle $Y={\cal O}\oplus {\cal T}$,
with ${\cal T}$ a line bundle over $B_2$ and ${\cal O}(1)$ a line bundle on 
the 
total space of ${\bf P}(Y)\rightarrow B_2$ which restricts on each ${\bf P}^1$ 
fiber to the typical line bundle over ${\bf P}^1$. Further let $a$, $b$ be 
homogeneous coordinates of the ${\bf P}^1$ bundle and think of $a$ and $b$ as 
sections, respectively, of ${\cal O}(1)$ and ${\cal O}(1)\otimes {\cal T}$ 
over $B_2$. If we set $r=c_1({\cal O}(1))$ and $t=c_1({\cal T})$ and 
$c_1({\cal O}\otimes {\cal T})=r+t$ then the cohomology ring of $B_3$ is 
generated over the cohomolgy ring of $B_2$ by the element $r$ with the 
relation r(r+t)=0, i.e. the divisors $a=0$ resp. $b=0$, which are dual to 
$r$ resp. $r+t$, do not intersect. The characteristic classes of $B_3$ 
can be derived from the total Chern class of $B_3$ given by
\beqa
c(B_3)=(1+c_1(B_2)+c_2(B_2))(1+r)(1+r+t)
\eeqa
one finds using ($c_i=\sigma^*(c_i(B_2))$)
\beqa  
c_1(B_3)&=& c_1+2r+t \\
c_2(B_3)&=& c_2+c_1t+2c_1r.
\eeqa
Furthermore, note that $B_2$ has to be rational which can be seen as follows:
Recall the arithmetic genus $p_a$ of $B_3$ has to be equal to one 
in order to satisfy the $SU(4)$ holonomy condition, otherwise there are non-
constant holomorphic differentials on $B_3$ which would pull back to X and 
thus destroy $SU(4)$ holonomy of X, i.e. $p_a=\int_{B_3}c_1(B_3)c_2(B_3)/24=1$
and using the expressions for $c_1(B_3)$ and $c_2(B_3)$ one finds  
$p_a=\frac{1}{12}\int_{B_2}c_1^2+c_2=1$ saying that $B_2$ is rational.

Further, it is known that for elliptic fibered Calabi-Yau threefolds $X_3$
the base $B_2$ has at worst log-terminal\footnote{A variety $X$ is said to
have at worst {\it log-terminal} singularities, if the following conditions 
are satisfied:

(i) $K_X$ is an element of $Div(X)\otimes {\bf Q}$ 
    (i.e.$X$ is {\bf Q}-Gorenstein) 

(ii) for a resolution of singularities $\rho:Y\rightarrow X$ such that the 
exceptional locus of $\rho$ is a divisor whose irreducible components $D_1...
D_r$ are smooth divisors with only normal crossings, one has 
$K_Y=\rho^* K_X+\sum_{i=1}^r a_iD_i$ with $a_i>-1$ (resp. $a_i\geq 0$) for
all $i$, where $D_i$ runs over all irreducible components of $D$ 
\cite{gra1},\cite{gra2},\cite{gra3}.} singularities. In particular one 
can consider \cite{V}
$B_2$'s which, after resolving of singularities, are either surfaces
whose canonical bundle is trivial, i.e. $K3$ surface, Enriques surface, 
a hyperelliptic surface, a torus, or they are $F_n$, $dP_k$, 
$\frac{T^2\times {\bf P^1}}{G}$. Here $F_n$ is the Hirzebruch surface, a 
${\bf P^1}$ bundle over ${\bf P^1}$ resp. the del Pezzo surface, get by
blow up of ${\bf P^2}$ in $k=1,...,k$ points.

Finally, let us recall that the expessions for the Euler numbers of 
$X_3$ and $X_4$ are given by \cite{SVW},\cite{klemm}
\beqa
\chi(X_3)&=& 2(h^{1,1}-h^{2,1})\\
\chi(X_4)&=& 6(8+h^{1,1}+h^{3,1}-h^{2,1}).
\eeqa 
All computations done so far are valid for elliptic fibrations which have a
section and can be described by a smooth Weierstrass model.

\newpage
%------------------------------------------------------------------------------
\subsection{Vector bundles on elliptic fibrations}
%------------------------------------------------------------------------------
To compactify the heterotic string on an elliptic 
Calabi-Yau threefold $Z$, one 
has to specify a vector bundle $V$ over $Z$, which breaks part of the 
$E_8\times E_8$ gauge symmetry. Friedman, Morgan and Witten have extensively 
studied the construction of vector bundles over $Z$ using the parabolic 
and spectral cover approach for $V$ \cite{FMW}, \cite{FMW2}, \cite{FMW3}. 

In the parabolic approach they considered a component of the moduli space of 
$SU(n)$ bundles (understood as rank $n$ complex vector bundles) which are 
$\tau$-invariant, i.e. the involution of the elliptic fibration 
$Z\rightarrow B$ lifts up to $V$. This condition 
could be implemented for $SU(n)$ bundles with $n$ even. 

In the spectral cover approach they considered  $V=SU(n)$ bundles 
with $n$ arbitrary \cite{FMW}. These bundles possess an additional 
degree of freedom since one has the possibility to 'twist' with a line 
bundle ${\cal N}$ on the spectral cover $C$, which leads to a multi-component 
structure of the moduli space of such bundles. Further, they did not require 
any $\tau$-invariance of $V$ in the construction, moreover, they 
found $\tau$-invariance for a certain class of these bundles which
have no additional twists. In particular, they found the same modulo conditions
(see below) as in the parabolic approach, in order to get $\tau$-invariance. 
Further, their $\tau$-invariant bundles have vanishing third Chern-class. 

In the following section we will adopt these approaches. In contrast, 
we will not work at the $\tau$-invariant point, we will construct \cite{A1}
$SU(n)$ vector bundles using the parabolic approach which have $n$ odd.
We compare our bundles with the spectral cover approach and we find that 
for a certain 'twist' of the line bundle ${\cal N}$, 
both approaches agree. In particular the bundles have a non-vanishing 
third Chern-class which leads to a non-vanishing net number of chiral 
fermions in heterotic string compactifications. 
 
%------------------------------------------------------------------------------
\subsubsection{Bundles from parabolics}
%------------------------------------------------------------------------------

In the parabolic approach one starts with an unstable
bundle on a single elliptic curve $E$ which is given by{\footnote{
here $W_k$ can be defined inductively as unique non-split extension
$0\rightarrow {\cal O}\rightarrow W_{k+1} \rightarrow W_{k}\rightarrow 0$
(c.f.\cite{FMW})}}
\beqa
V=W_k\oplus W_{n-k}^*
\eeqa
this has the property that it can be deformed by an element $\alpha\in H^1(E,
W^*_k\otimes W^*_{n-k})$ to a (semi) stable bundle $V'$ over $E$ which fits in
the exact sequence 
\beqa
0\rightarrow W^*_{n-k}\rightarrow V'\rightarrow W_k\rightarrow 0
\eeqa
Now, to get a global version of this construction one is interested in 
an unstable bundle over $Z$, which 
reduces on every fibre of $\pi:Z\rightarrow B$ to the unstable 
bundle over $E$ and can be deformed to a stable bundle over
$Z$. Since the basic bulding blocks were on each fiber $W_k$ with 
$W_1={\cal O}(p)$, one has to replace them by their global versions.  
So, global one replaces $W_1={\cal O}(p)$ by $W_1={\cal O}(\sigma)$ 
and $W_k$ can be defined inductively by an exact sequence
\beqa
0\rightarrow {\L}^{n-1}\rightarrow W_k \rightarrow W_{k-1}\rightarrow 0
\eeqa
with ${\cal L}=K_B^{-1}$.
Further, one has globally the posibility to twist by additional data coming
from the base $B$ and so one can write \cite{FMW} for $V=SU(n)$ 
\beqa
V=W_k\otimes{\cal M}\oplus W^*_{n-k}\otimes{\cal M}^{\prime}.
\eeqa
Note that the unstable bundle can be deformed to a stable one by an element
in $H^0(B,R^1\pi_*(ad(V)))$ since the Leray spectral sequence degenerates
to an exact sequence \cite{FMW}. Further, note that for the computation
of characteristic classes we can use the unstable bundle because the topology
of the bundle is invariant under deformations.  
%------------------------------------------------------------------------------

%------------------------------------------------------------------------------
Now, let us start with the unstable $G=SU(n)$ bundle $V$
where ${\cal M}, {\cal M}^{\prime}$ are line bundles over $B$ which are 
constrained so that $V$ has trivial determinant, i.e. ${\cal M}^k\otimes({\cal M}^{\prime})^{n-k}\otimes{\L}^{-\frac{1}{2}(n-1)(n-2k)}\cong{\cal O}$.
Further, $W_k$ and $W^*_{n-k}$ are defined as
\beqa
W_k=\bigoplus_{a=0}^{k-1}{\L}^a, \ \ \ \
W^*_{n-k}=\bigoplus_{b=0}^{n-k-1}{\L}^b
\eeqa
where we have set ${\L}^0={\cal O}(\sigma)$ and 
${\L}^{-0}={\cal O}(\sigma)^{-1}$ with $\L$ being a line bundle on $B$. 
The total Chern class of $V$ can then be written as
\beqa
c(V)=\prod_{a=0}^{k-1}(1+c_1(\L^a)+c_1(\M))
     \prod_{b=0}^{n-k-1}(1+c_1(\L^{-b})+c_1(\M^{\prime})).
\label{total}
\eeqa
In the following we will discuss the two cases: $n$ is even and $n$ is odd.
First let us review the case that $n$ is even which was considered
in \cite{FMW}, then we will extend to the case $n$ is odd and compare both
cases with the spectral cover construction of $V$.\\ \\  
{\underline{$n$ even}}\\ \\
In this case one can choose $k=\frac{n}{2}$ which
restricts one to take ${\cal M}^{\prime}={\cal M}^{-1}$ in order to obey
trivial determinant of $V$. The advantage of taking $k=\frac{n}{2}$ is that
the condition of $\tau$ invariance of $V$ is easily implemented.  
Note that $\tau$ operates on $V$ as $\tau^*V=V^*$, i.e. $k\rightarrow n-k$.
Now expansion of the total Chern class of $V$ 
immediatly leads to $c_1(V)=0$ and $c_3(V)=0$. Further setting 
$\sigma=c_1({\cal O}(\sigma))$ and $\eta=-2c_1(\M)+c_1(\L)$
respectively using the fact that $\sigma^2=-\sigma c_1(\L)$, one obtains  
for the second Chern class 
\beqa
c_2(V)=\eta\sigma-\frac{1}{24}c_1(\L)^2(n^3-n)-\frac{n}{8}\eta(\eta-nc_1(\L)).
\eeqa    
%\beqa
%c_2(V)=-2\sigma c_1(\M)-\sigma^2+\left(\frac{-n^3+3n^2-2n}{24}
%\right)c_1(\L)^2+
%\left(\frac{2n-n^2}{4}\right)c_1(\M)c_1(\L)\\-\frac{n}{2}c_1(\M)^2
%\nonumber
%\eeqa
Moreover, from $c_1(\M)=-\frac{1}{2}(\eta-c_1(\L))$ one gets the congruence 
relation for $\eta$:
\beqa
\eta\equiv c_1(\L) \ \ (mod\ \ 2)
\eeqa
Thus $c_2(V)$ is uniquely determined in terms of $\eta$ and the elliptic
Calabi-Yau manifold $Z$. In particular one has
\beqa
\eta=\pi_*(c_2(V)).
\eeqa
Now let us turn to the case that $n$ is odd! \\ \\
\underline{$n$ odd}\\ \\
Let us first specify \cite{A1} our unstable bundle $V$. Actually we have $n$ 
different choices to do this which depend on the choice of the integer $k$
in the range $1\le k \le n$. We will choose 
$k=\frac{n+1}{2}$ and the line bundles ${\cal M}={\cal S}^{\frac{-n+1}{2}}$ 
and ${\cal M}'={\cal S}^{\frac{n+1}{2}}\otimes {\L}^{-1}$ which will be
presently shown to be  
appropriate in order to compare it with results 
obtained from spectral covers for $V$. Therefore
we can write for our unstable bundle $V$
\beqa
V=W_k\otimes {\cal S}^{-n+k}\oplus W^*_{n-k}\otimes {\cal S}^k\otimes {\L}^{-1}
\eeqa
which has trivial determinant. Using the above relation for the total 
Chern-class of $V$ and setting again $\sigma=c_1({\cal O}(\sigma))$ 
and ${\sigma}^2= -\sigma c_1({\L})$, and
\beqa
\eta= n c_1({\cal S})
\eeqa
we will find for the characteristic classes of $V$ 
\beqa
c_1(V)&=&0\\
c_2(V)&=&\eta\sigma-\frac{1}{24}c_1(\L)^2(n^3-n)-
       \frac{n}{8}\eta(\eta-nc_1(\L))+\frac{1}{8n}\eta(\eta-nc_1(\L))\label{1}
\\
c_3(V)&=&\frac{1}{n}\sigma\eta(\eta-nc_1(\L))
\eeqa
So, we are restricted to bundles $V$ with $\eta=\pi_*(c_2(V))$ 
divisible by $n$. 

Now, integration of $c_3$ over $Z$ can be accomplished by first integrating
over the fibers of $Z\rightarrow B$ and then integrating over the base. 
Further, using the fact that the section $\sigma$ intersects the fiber $F$
in one point $\sigma\cdot F=1$ and $r=3\sigma$ where $r$ is the cohomology
class dual to the vanishing of the section of the line bundle ${\cal O}(1)$
(defined on the total space of the Weierstrass model of $Z$), we get
\beqa
\int_Z c_3(V)=\int_B \frac{1}{n}\eta(\eta-nc_1(\L)).
\eeqa
This leads to the net number $N_{gen}$ of generations in a heterotic string 
compactification
\beqa
N_{gen}=n_{gen}-{\bar n}_{gen}=\int_B \frac{1}{2n}\eta(\eta-nc_1(\L))
\eeqa
and thus, if we fix the elliptic manifold $Z$, so that the section $\sigma$
and $c_1(\L)$ are fixed, $N_{gen}$ is uniquely determined for a choice of 
$\eta$. Here we have to note that in order to determine 
separately the number of generations $n_{gen}$ respectively 
antigenerations ${\bar n}_{gen}$ instead
of just their difference, we have in addition to compute the dimension of
$H^1(Z,V)$. This can be done by using the Leray spectral sequence. \\

In order to compare our results with those obtained from the spectral 
cover construction let us review some facts of the setup \cite{FMW}.

%------------------------------------------------------------------------------
\subsubsection{Bundles from spectral covers}
%------------------------------------------------------------------------------

We are interested in $SU(n)$ vector bundles $V$ over the elliptically fibered 
Calabi-Yau threefold $\pi:Z\rightarrow B$ with a section $\sigma$ of
$K_B=\L^{-1}$ and let $Z$ be represented by its Weierstrass model, which
embeds $Z$ in a ${\bf P^2}$ bundle over $B$, as above, we have
$y^2=x^3-g_2x-g_3$. In the spectral cover approach the vector bundle over $Z$ 
is described by a pair ($C,{\cal N}$), more precisely $V$ can be reconstructed
from this pair, where $C$ is the spectral cover and 
${\cal N}$ an arbitrary line bundle over $C$. The spectral cover $C$ is 
given by the vanishing of a section $s$ of ${\cal O}(\sigma)^n\otimes 
{\cal M}$ with $\M$ being an arbitrary line bundle over $B$ of $c_1(\M)=\eta'$.
The locus $s=0$ for the section is given for $n$ even by 
\beqa
s=a_0z^n+a_2z^{n-2}x+a_3z^{n-3}y+...+a_nx^{n/2}
\eeqa
(resp. the last term is $x^{(n-3)/2}y$ for n odd)\footnote {here 
$a_{r}\in\Gamma(B,\M\otimes\L^{-r})$, $a_0$ is a section of $\M$ and $x$, $y$ 
sections of $\L^2$ resp. $\L^3$ in the Weierstrass model (c.f.\cite{FMW})}
which determines a hypersurface in $Z$.
If we fix a point in $B$ then the Weierstrass equation and the locus $s=0$
have $n$ solutions and so $C$ is an $n$-fold ramified cover of $B$. 
Further, one recovers $V$ from $C$ by choosing the Poincare line bundle 
${\cal P}_B$ over the fiber product $Z\times_B Z$ 
and considers its restriction to the subspace $Y=C\times_B Z$ of $Z\times_B Z$.
Taking the projection to the second factor of the fiber product 
$\pi_2:Y\rightarrow Z$, $V$ can be constructed via push forward of the 
Poincare bundle on $Y$ to $Z$. Further, taking into account that ${\cal P}_B$
can be twisted by the pull back (via $\pi_1$) of the line bundle ${\cal N}$, 
so one has \cite{FMW}
\beqa
V=\pi_{2*}({\cal N}\otimes {\cal P}_B)
\eeqa
It is important to note that ${\cal N}$ is not completely arbitrary, it is
restricted through the condition of vanishing first Chern-class of $V$. 
Since the Poincare line bundle becomes trivial when restricted to $\sigma$,
one can use Grothendieck-Riemann-Roch for the projection 
$\pi:C\times_B \sigma= C\rightarrow B$ and gets
\beqa
\pi_*(e^{c_1({\cal N})}Td(C))=ch(V)Td(B)
\eeqa
and with the condition $c_1(V)=0$ one has 
\beqa
c_1({\cal N})=-\frac{1}{2}(c_1(C)-\pi_*c_1(B))+\gamma
=\frac{1}{2}(n\sigma+\eta'+c_1(\L))+\gamma
\eeqa
here $\gamma\in H^{1,1}(C,{\bf Z})$ with $\pi_*\gamma=0\in H^{1,1}(B,{\bf Z})$.
In particular if one denotes by $K_B$ and $K_C$ the canonical bundles of $B$ 
and $C$ then one has \cite{FMW}
\beqa
{\cal N}=K_C^{1/2}\otimes K_B^{-1/2}\otimes ({\cal O}(\sigma)^n\otimes\M^{-1}
\otimes \L^n)^{\lambda}
\eeqa
from which one learns that $\gamma=\lambda(n\sigma-\eta'+nc_1(\L))$. 

The second Chern class of $V$ is computed in \cite{FMW} and given by
\beqa
c_2(V)=\eta'\sigma-\frac{1}{24}c_1(\L)^2(n^3-n)-\frac{n}{8}\eta'(\eta'
-nc_1(\L))-\frac{1}{2}\pi_*(\gamma^2)
\eeqa
where the last term reflects the fact that one can twist with a line bundle
${\cal N}$ on the spectral cover, one has
\beqa
\pi_*(\gamma^2)=-\lambda^2n\eta'(\eta'-nc_1(\L)).
\eeqa
For later use we will $c_2(V)$ denote (following \cite{FMW}) by 
\beqa
c_2(V)=\sigma \eta +\omega
\eeqa
where $\omega \in H^4(B)$, i.e. $\omega=c_2(V|_B)$ and
\beqa
\omega=-[\frac{n^3-n}{6}\frac{c_1^2}{4}+\frac{n}{8}\eta(\eta-nc_1)+
\frac{1}{2}\pi_*(\gamma^2)]
\label{om}
\eeqa
Now let us compare this with 
the parabolic computation we have done! 
Note that, $\lambda\neq 0$ (more precisely $\gamma\neq 0$) measures the 
deviation from $\tau$-invariance for bundles in the spectral cover 
construction. We will start
with the two cases $\lambda=0$, $\lambda=\frac{1}{2}$ mentioned in 
\cite{FMW} and then consider our case $\lambda=\frac{1}{2n}$.\\
\\
In case that $n$ is even, it was shown \cite{FMW} that to achieve 
$\tau$ invariance in the spectral cover approach for $V$, one must define
${\cal N}$ in the above sense with $\gamma=0$, i.e. $\lambda=0$. In 
particular the existence of an isomorphism ${\cal N}^2=K_C\otimes
K_B^{-1}$ was shown. Further, it was 
shown that there are the same mod two coditions 
for $\eta'$ and $n$ in the spectral cover approach for $\gamma=0$ and in 
particular that also $\eta'=\pi_*(c_2(V))$ and therefore one is lead to the
identification $\eta'=\eta$. \\ 
\\
In case that $n$ is even and $\lambda=\frac{1}{2}$ 
the last two terms in combine in (\ref{om}) the only
general elements of $H^{1,1}(C,{\bf Z})$ are 
$\sigma|_C$ and $\pi^* \beta$ (for $\beta \in H^{1,1}(B,{\bf Z})$), 
which have because of $C=n\sigma + \pi ^*\pi_* c_2V$
the relation $\pi_* (\sigma|C)=\pi_* \sigma(n\sigma+\pi^*\eta)=
\pi_* \sigma(-nc_1+\pi^*\eta)=\eta-nc_1$;
so $\gamma=\lambda (n\sigma-\pi^*(\eta-nc_1))$ (with $\lambda$ 
possibly half-integral) and $\pi_*(\gamma^2)=-\lambda^2 n\eta(\eta-nc_1)$;
so for $\lambda=1/2$ the term would disappear.\\
\\
Now in our case, if $n$ is odd, we can identify the last term in ({\ref{1}})
with $\frac{\pi_*(\gamma^2)}{2}$ if we choose $\lambda=\frac{1}{2n}$, i.e. 
the parabolic approach for $n$ odd agrees with the spectral cover approach
if we choose the twisting line bundle ${\cal N}$ appropriate on the 
spectral cover. Furthermore, if we use 
\beqa
c_1({\cal N})=\frac{1}{2}(n\sigma+\eta'+c_1(\L))+\gamma=\frac{(n+1)}{2}\sigma
+c_1(\L)+\frac{(n-1)}{2} \frac{\eta'}{n}
\eeqa
which is well defined for $n$ odd and since we can choose 
${\cal M}={\cal S}^n$ we are left with ${\eta}'=n c_1({\cal S})$ and so 
with $\lambda=\frac{1}{2n}$ we have the same conditions 
for $\eta'$ and $n$ as we had in the parabolic approach.\\
\\
{\bf{Discussion:}} We have constructed a class of 
$SU(n)$ vector bundles, with $n$ odd, 
in the parabolic bundle construction, which have a 
$\eta\equiv 0 ({\rm{mod}}\ \ n)$ condition,
in contrast to the bundles, which have $n$ even and a
$\eta\equiv c_1(\L) ({\rm{mod}}\ \ 2$)) condition.  
For $n$ even, the bundles in the parabolic construction are restricted 
to the $\tau$-invariant bundles
in the spectral cover construction, given at $\lambda=0$. Our bundles have no 
$\tau$-invariance and being restricted to bundles of
$\lambda=\frac{1}{2n}$, in the  
spectral cover construction.
\\ \\
{\bf{Remark:}} In \cite{CP}
the computation of $c_3(V)$ was performed in the spectral cover approach,
it is 
\beqa
c_3(V)=2\lambda \eta(\eta-nc_1).
\eeqa
and so in nice agreement with our computation of $c_3(V)$ in the 
parabolic approach at our special point $\lambda=\frac{1}{2n}$! 

\newpage

%------------------------------------------------------------------------------
\subsection{Bundle moduli}
%------------------------------------------------------------------------------

After describing the construction of vector bundles over an elliptic
fibered Calabi-Yau manifold and computing the Chern classes of $V$ we 
have now to determine an additional invariant characterizing our bundle,
the dimension of $H^1(Z,{\rm{End}}(V))$. This dimension will later 
play an important role, since the number of bundle moduli contributes to 
the massless spectrum in the heterotic string compactification. 

In the first section of this chapter, we will review the computation 
for $Z=K3$ originally performed for the general case in \cite{koba}.   
Then we turn to the threefold case and count the moduli $h^1(Z,{\rm{ad}}(V))$ 
for $V=E_8$ by applying \cite{ACL} an index computation, which was first 
used in this specific form by Friedman, Morgan and Witten \cite{FMW} for 
$V=SU(n)$. 

%------------------------------------------------------------------------------
\subsubsection{Bundles over surfaces}
%------------------------------------------------------------------------------

Let us determine the number of bundle moduli in the case that $Z$ is 
a $K3$ surface. We start with the Hirzebruch-Riemann-Roch formula 
\beqa
\chi(Z,{\rm{End}}(V))&=&\int_{Z}td(Z)ch({\rm{End}}(V))
               =\int_{Z}td(Z)ch(V)ch(V^*)\nonumber\\
              &=&\int_{Z}-2nc_2(V)+\frac{n^2}{12}(c_1(Z)^2+c_2(Z))
\eeqa
Now, the Riemann-Roch formula gives
\beqa
1-h^{0,1}(Z)+h^{0,2}(Z)=\frac{1}{12}\int_{Z}c_1(Z)^2+c_2(Z).
\eeqa
Further, with
\beqa
\chi(Z,{\rm{End}}(V))=\sum_{i=0}^{2}(-1)^ih^i(Z,{\rm{End}}(V))
\eeqa
we find for the dimension of the moduli space setting $h^0(Z,{\rm{End(V)}})=0$ 
and $h^2(Z,{\rm{End(V)}})=0$ 
(in order to get a smooth moduli space \cite{koba}, 
\cite{hyprecht}) we get
\beqa
h^1(Z,{\rm{End(V)}})=2nc_2(V)+n^2h^{0,1}(Z)-(n^2-1)(1+h^{0,2})
\eeqa
using the fact that $h^{0,1}(Z)=0$ and $h^{0,2}(Z)=1$ for $K3$
we get
\beqa
h^{1}(Z,{\rm{End(V)}})=2(nc_2(V)+1-n^2)
\eeqa

%------------------------------------------------------------------------------
\subsubsection{Bundles over elliptic threefolds}
%------------------------------------------------------------------------------

Let us recall the
setup of the index-computation in \cite{FMW}
concerning the contribution of the bundle moduli in the case of 
elliptic Calabi-Yau threefolds. 

As the usual quantity
suitable for index-computation $\sum_{i=o}^3 (-1)^i h^i(Z,{\rm{End}}(V))$ 
vanishes
by Serre duality, one has to introduce a further twist to compute a
character-valued index. Now because of the elliptic fibration structure
one has on $Z$ the involution $\tau$ coming from the "sign-flip" in the fibers
and we furthermore assume that at least at some point, at the $\tau$-invariant 
point, in moduli space the 
symmetry can be lifted to an action on the bundle. 
In particular, the action of $\tau$ lifts to an action on the adjoint bundle
${\rm{ad}}(V)$, which are the traceless endomorphisms of ${\rm{End}}(V)$.
If one projects onto
the $\tau$-invariant part of the index problem one has
\beqa
I=-\sum_{i=o}^3 (-1)^i Tr_{H^i(Z,{\rm{ad}}(V))}\frac{1+\tau}{2}= -\sum_{i=o}^3 
(-1)^i h^i(Z,{\rm{ad}}(V))_e
\eeqa
where the subscript "e" (resp. "o") indicates the even (resp. odd) subspaces
of $H^i(Z,{\rm{ad}}(V))$ under $\tau$ and we used 
\beqa
Tr_{H^i(Z,{\rm{ad}}(V))}\frac{1+\tau}{2}= h^i(Z,{\rm{ad}}(V))_e.
\eeqa
The character-valued index simplifies by the vanishing of the ordinary index 
to
\begin{eqnarray}
I=-\frac{1}{2}\sum_{i=o}^3 (-1)^i Tr_{H^i(Z,{\rm{ad}}(V))} \tau .
\end{eqnarray}
If we have an unbroken gauge group $H$, which is the commutator of the group
$G$ of $V$, then $I=n_e-n_o$ has to be corrected by $h^0_e-h^0_o$
denoting by $n_{e/o}$  the number $h^1(Z,{\rm{ad}}(V))_{e/o}$ of 
massless even/odd
chiral superfields and by $h^0_{e/o}$ the number of unbroken gauge group
generators even/odd under $\tau$ \cite{FMW}. So one has for the 
number of bundle moduli 
\beqa
m_{bun}=h^1(Z,{\rm{ad(V)}})=n_e+n_o=I+2n_o
\eeqa
Using a fixed point theorem, as it was extensively done in \cite{FMW}, one
can effectively compute the character-valued index $I=n_e-n_o$ of bundle
moduli 
\beqa
I=rk-\sum_j\int_{U_j}c_2(V)
\eeqa
where the $U_j$ denotes the two fixed point sets which we want to describe
explicitly in the following. Therefore recall that $Z$ is described by a
Weierstrass equation and $\tau$ is the transformation $y\rightarrow -y$
which leaves the other coordinates fixed. At a fixed point the coordinates
$x$, $y$ and $z$ are left fixed up to overall rescaling. From the 
Weierstrass equation 
\be
y^2z = x^3+g_2xz^2+g_3z^3
\eq
we see that there are two components of the fixed point set. 

The first component, denoted by $U_1$, is given in homogeneous 
coordinates by $(x,y,z)=(0,1,0)$ which is the standard section 
$\sigma$ of $Z\rightarrow B$ and thus isomorphic to a copy of $B$ and 
the cohomology class of $U_1$ is the class of the section.

The second component, denoted by $U_2$, is given by the triple cover
of $B$ defined by
\be
0 = x^3+g_2xz^2+g_3z^3
\eq
which embeds $U_2$ in a ${\bf P}^1$ bundle $W\rightarrow B$. As already
noted above, we can think of $x$ and $z$ as sections of ${\cal O}(1)\otimes 
K_B^{-2}$ and  ${\cal O}(1)$ with first Chern classes $r+2c_1$ and $r$ 
respectively. The cohomology ring of $W$ is generated over the 
cohomology ring of $B$ by the element $r$ with relation $r(r+2c_1)=0$, i.e.
$x$ and $z$ have no common zeros or in other words the divisors dual to
$x=0$ and $z=0$ do not intersect.

Now $U_2$ is defined by the vanishing of a section of ${\cal O}(1)^3
\otimes K_B^{-6}$ which is a line bundle over $W$ with first Chern class
$3r+6c_1$ and multiplication by $3(r+2c_1)$ can be understood as restricting 
$W$ to $U_2$. The relation $r(r+2c_1)=0$ can be simplified in the cohomology
ring of $U_2$ to $r=0$. 

With this in mind, we are able to compute the index $I$ using the identity
(c.f.\cite{ACL}) 
\beqa
\int_{U_i}c_2(V)=\int_{B}c_2(V)|_{U_1}+3\int_{B}c_2(V)|_{U_2}
\eeqa
where 3 appears since $U_2$ is a triple cover of $B$ (and for $E_8$ bundles 
replace $c_2(V)$ by $\lambda(V)$). Next let us determine
the number of bundle moduli for $E_8$ and $SU(n)$ vector bundle!\\ 
{\bf{Note:}} In \cite{ACL} we performed the derivation in a slightly different
way, using the nonperturbative anomaly cancellation condition and restricting
$c_2(Z|_{U_i})+[W]|_{U_i}$, to the fixed point set. \\
{\bf{Bundle moduli for $V=E_8$, $SU(n)$}:}\\ 
The fundamental characteristic class $\lambda(V)$\footnote{
$\lambda(V)=c_2(V)/60$} of an $E_8$ vector bundle is given by \cite{FMW}
\beqa
\lambda(V)=\eta\sigma-15\eta^2+135\eta c_1-310c_1^2
\eeqa
which restricts to the fixed point set as
\beqa
\int_{B}\lambda(V)|_{U_1}&=&\int_{B}(-\eta c_1-15\eta^2+135\eta c_1-310c_1^2)\\
3\int_{B}\lambda(V)|_{U_2}&=&\int_{B}(-45\eta^2+405\eta c_1-930c_1^2)
\eeqa
and we will find for the index of the $E_8$ bundle using $\lambda(V)$ 
\cite{ACL}
\beqa
I=8-4(\lambda(V)-\eta\sigma)+\eta c_1.
\eeqa
Further the fundamental characteristic class of an $SU(n)$ bundle is given by
\beqa
c_2(V)=\eta\sigma+\omega
\eeqa
restriction to the fixed point set leads to
\beqa
\int_{B}c_2(V)|_{U_1}&=&\int_{B}(-\eta c_1+\omega)\\
3\int_{B}c_2(V)|_{U_2}&=&\int_{B}(3\omega)
\eeqa
and the index of the $SU(n)$ bundle is given by \cite{FMW}
\beqa
I=n-1-4(c_2(V)-\eta\sigma)+\eta c_1 \label{indi}.
\eeqa
Let us also consider the case that $V=TZ$ the tangential bundle to Z \cite{A2}!
Therefore recall the second Chern-class of $Z$ was given by
\beqa
c_2(Z)=c_2+11c_1^2+12\sigma c_1
\eeqa
so, we find for the restriction to the fixed point set
\beqa
\int_{B}c_2(Z)|_{U_1}&=&\int_{B}(c_2-c_1^2)\\
3\int_{B}c_2(Z)|_{U_2}&=&\int_{B}(3c_2+33c_1^2)
\eeqa
thus the index for the $SU(3)$ bundle is given by
\beqa
I&=&2-4c_2-32c_1^2\nonumber\\
&=& -46-28c_1^2.
\eeqa

\newpage

%------------------------------------------------------------------------------
\section{F-Theory}  
%------------------------------------------------------------------------------
\resetcounter
%------------------------------------------------------------------------------
\subsection{The framework}
%------------------------------------------------------------------------------
As already mentioned in the introduction, F-theory can be considered \cite{V}
as a twelve-dimensional theory underlying the conjectured $SL(2,{\bf Z})$ 
duality symmetry of type IIB in ten-dimensions. 

Therefore recall, type IIB in ten-dimensions contains the Neveu-Schwarz-Neveu-
Schwarz (NS-NS) sector with the bosonic fields $g_{MN},B_{MN}^A,\phi$ and the
Ramond-Ramond (R-R) sector with the bosonic fields $B_{MN}^P,{\tilde\phi},
A_{MNPQ}^+$, where $\phi$ is the dilaton and ${\tilde\phi}$ the axion which
combine into $\tau={\tilde\phi}+i\exp(-\phi)$. In particular, the equations 
of motion of low energy type IIB supergravity are covariant, and invariant 
under $SL(2,{\bf R})$ transformations \cite{GS143},\cite{SW301},\cite{S269}.
It has been conjectured that $SL(2,{\bf R})$ is broken to its maximal compact
subgroup $SL(2,{\bf Z})$ \cite{HT}. The metric and the 
four-form potential are invariant
under $SL(2,{\bf Z})$, further, $\tau$ transfroms as 
$\tau\rightarrow \frac{a\tau+b}{c\tau+d}$ where $(a,b,c,d)\in{\bf Z}$ 
with $ad-cb=1$ and the antisymmetric tensor fields get exchanged.
Thus $\tau$ transforms in the same way as the complex modulus of the torus. 
Now, in pure perturbative type IIB compactifications $\tau$ is constant. Here
the F-theory starts! Compactifications of F-theory to lower dimensions can be 
formulated as type IIB theory on a manifold $B$ with varying $\tau$ over 
$B$ \cite{V}. So, compactification of F-theory on a manifold $X$ which
admits an eliptic fibration over $B$ can be considered as type IIB on $B$ with
$\tau=\tau(z)$ and $z\in B$. Regarding F-theory in this way avoids the 
difficulties in formulating a consistent twelve-dimensional theory.    
Now $X$ can be represented by its Weierstrass model 
\beqa
y^2=x^3+xf(z_i)+g(z_i)
\eeqa
with $f,g$ functions on the base $B$ of degree 8 resp. 12. The elliptic fibre 
degenerates when $\Delta=4f^3+27g^2$ vanishes. The class of the discriminant
(as given above) is $D=-12K_B$, localized at codimension one in the 
base $B$. For example, if $X=K3$ the elliptic fibre degenerates over 24 points
in $B={\bf P}^1$, moreover, writing $\Delta=c\prod_{i=1}^{24}(z-z_i)$ and  
noting that $\tau\sim \frac{1}{2\pi i}\log(z-z_i)$, one finds that the $\tau$
modulus becomes singular if $z\rightarrow z_i$. If we go around $z=z_i$, we 
have $\tau\rightarrow \tau+1$ and this implies a discrete shift of 
$\tilde{\phi}\rightarrow \tilde{\phi}+1$, which has no explanation in
perturbative type IIB compactification, in particular, this shift signals 
the presence of a magnetically charged 7-brane in $z=z_i$ filling the 
uncompactified spac-time \cite{V} (also c.f. \cite{cosmic}). 

The most simple degenerations of the elliptic fibre are of type I$_1$ 
(in the Kodaira classification) and the corresponding Weierstrass model
is smooth. Unbroken gauge symmetry can be obtained in F-theory, if the 
elliptic fibre admits a section of higher singularities then I$_1$, say
a section of $E_6$ singularities, would then correspond in the conjectured
het/F-theory duality to an unbroken $E_6$ space-time gauge group. 
Furthermore, on the compact part of the world-volume of the 7-brane one can 
turn on a non-trivial gauge field background with a nonzero instanton number
\cite{W460}. The presence of such a background would then further break the 
gauge symmetry.

%------------------------------------------------------------------------------
\subsection{The constraints}
%------------------------------------------------------------------------------
In order to obtain consistent F-theory compactifications to four-dimensions
on Calabi-Yau fourfold $X$, the necessity of turning on a
number $n_3$ of space-time filling three-branes for tadpole cancellation 
was established \cite{SVW}. This is motivated by the fact that 
compactifications of
the type IIA string on $X$ are destabilized at one loop by 
$\int B\wedge I_8$, where $B$ is the NS-NS two-form which couples 
to the string and $I_8$ a linear 
combination of the Pontryagin classes $p_2$ and $p_3$ \cite{VW}. 
So, compactifications of the type IIA string to two-dimensions leads to
a tadpole term $\int B$ which is proportional to the Euler characteristic
of $X$. 
Similar, in
M-theory compactifications to three-dimensions on $X$ arises a term 
$\int C\wedge I_8$ with $C$ being the M-theory 
three-form, and integration over $X$ then leads to the tadpole term
which is proportional to $\chi(X)$, 
and couples to the 2-brane \cite{DLM},\cite{BB}. Now, as M-theory 
compactified to three-dimensions on $X$ is expected to be related to a 
F-theory compactification to four-dimensions
\cite{V}, one is lead to expect a term 
$\int A$ with $A$ now being the
R-R four form potential, which couples to the three-brane in F-theory 
\cite{SVW}. Taking into account the proportionality 
constant \cite{Wphase}, one finds $\frac{\chi(X)}{24}=n_3$ in F-theory
(or $=n_2$ in M-thoery resp. strings in type IIA theory) \cite{SVW}.    

Furthermore, it was shown that the tadpole in M-theory will be corrected
by a classical term $C\wedge dC\wedge dC$, which appears, if $C$ gets
a background value on $X$ and thus leads to a contribution $\int dC\wedge dC$
to the tadpole \cite{BB}. Also, it has been shown \cite{W4flux} that one 
has as quantization law for the four-form field strength $G$
of $C$ (the four-flux)
the modified integrality condition $G=\frac{dC}{2\pi}=\frac{c_2}{2}+
\alpha$ with $\alpha\in
H^4(X,{\bf Z})$ where $\alpha$ has to satisfy the bound \cite{DC}
$-120-\frac{\chi(X)}{12}\le \alpha^2+\alpha c_2\le -120$, in order to keep
the wanted amount of supersymmetry in a consistent compactification.   
Also, the presence of a non-trivial instanton background can contribute 
to the anomaly \cite{BJPS}. 
Including all, tadpole cancellation and the four-flux condition and non-trivial
instanton background, one finds
\cite{DM}
\beqa
\frac{\chi(X)}{24}=n_3+\frac{1}{2}\int G\wedge G+\sum_{j} 
\int_{\Delta_j} c_2(E_j)
\eeqa
for consistent $N=1$ F-theory compactifications on $X$ to four-dimensions where
$\int_{\Delta_j} c_2(E_j)=k_j$ are the instanton numbers of possible background
gauge bundles $E_j$ inside the 7-brane \cite{BJPS} and $\Delta_j$ denotes the
discriminant component in $B_3$. 
Before going into details let us shortly recall how to establish
the heterotic/F-theory duality!
%------------------------------------------------------------------------------
\subsection{The dualities}
%------------------------------------------------------------------------------
The basic idea for establishing a duality 
between two theories in lower dimensions
is to use the {\it adiabatic principle} \cite{VW}. Therefore one has first 
to establish a duality between two theories, then one 
variies 'slowly' the parameters of these two theories over a common base space,
and one expects that the duality holds on the 
lower dimensional space \cite{V}. Furthermore, at those points where the 
adiabaticity breaks down, one expects new physics in the lower dimensional 
theory to come in. 

Let us be more precise! The heterotic string compactified on a 
$n-1$-dimensional elliptically fibered Calabi-Yau $Z\rightarrow B$ together
with a vector bundle $V$ on $Z$ is conjectured to be dual to F-theory 
compactified on a $n$-dimensional Calabi-Yau $X\rightarrow B$, fibered over 
the same base $B$ with elliptic $K3$ fibers. A duality between the two 
theories involves the comparision of the moduli spaces on both sides. 
In the following we will be interested in $n=2,3,4$. 
\\ \\    
Let us see how this works in various dimensions!

\newpage
%------------------------------------------------------------------------------
\section{Heterotic/F-Theory duality}
\resetcounter
%------------------------------------------------------------------------------
\subsection{8D Het/F-duality}
%------------------------------------------------------------------------------
\subsubsection{Heterotic string on $T^2$}
%------------------------------------------------------------------------------
Let us compactify the heterotic string on $T^2$. Let us represent $T^2$
by its Weierstrass equation, i.e. a cubic in ${\bf P}^2$
\beqa
y^2=x^3+g_2x+g_3
\eeqa
In addition one has to specify a vector bundle on $T^2$. This can be done 
by turning on Wilson lines on $T^2$. In particular, turning on 16 Wilson 
lines on $T^2$, breaks $E_8\times E_8$ to $U(1)^{16}$, otherwise, we are left
with the unbroken 10D gauge group, which involves a description of $E_8\times
E_8$ bundles on $T^2$. It has been shown \cite{FMW} that $E_8$ bundles 
on $Z$ can be described by embedding $Z$ in a rational elliptic surface, which
can be obtained by blowing up 9 points in ${\bf P}^2$. 

Now, on the level of parameter counting 
one gets 16 complex parameters coming from the Wilson lines, and additional
2 complex parameters from the complex structure modulus $U$ and the K\"ahler
modulus $T$ of $T^2$. These 18 parameters parametrize the moduli space
\cite{Na1},\cite{NSW}
\beqa
{\M}_{het}=SO(18,2;{\bf Z}){\backslash}SO(18,2){/}SO(18)\times SO(2)  
\eeqa
further, one has to take into account the heterotic coupling constant, which
is parametrized by a positive real number $\lambda^2$ \cite{V}, so we have
18 complex + 1 real parameters.

%------------------------------------------------------------------------------
\subsubsection{F-theory on $K3$}
%------------------------------------------------------------------------------
Now, in \cite{V} it was argued (on the level of parameter counting)
that the heterotic string on $T^2$ in the 
presence of Wilson lines is dual to F-theory compactified on $K3$ given by 
\beqa
y^2=x^3+g_2(z)x+g_3(z)
\eeqa
where the equation describes a hypersurface in a ${\bf P}^2$ bundle 
over ${\bf P}^1$. In particular one has $(9+13-3-1)=18$ parameters, where
$9+13$ coming from specifying $g_2$ and $g_3$, then mod out by $SL(2,{\bf C})$
action on ${\bf P}^1$ means just subtracting 3 and an additional 1 for 
overall rescalings. Furthermore, the remaining real parameter (the heterotic
coupling) can be identified with the size of the ${\bf P}^1$. 
\beqa
{\M}_{F}=SO(18,2;{\bf Z}){\backslash}SO(18,2){/}SO(18)\times SO(2)  
\eeqa

Further, it was argued \cite{MV1} that, in case of switching off Wilson lines,
the F-theory dual is given by the two-parameter family of $K3$'s
\beqa
y^2=x^3+\alpha z^4 x+(z^5+\beta z^6 +z^7)
\eeqa
with $E_8$ singularities at $z=0,\infty$. Therefore, it must exist a map that
relates the $T$ and $U$ modulus of the heterotic string to the two complex
structure moduli $\alpha$ and $\beta$. This map was made explicit in 
\cite{CCML} using the relationship between the $K3$ discriminat and the 
discriminant of the Calabi-Yau threefold $X_{1,1,2,8,12}(24)$ in the 
limit of large base ${\bf P}^1$.

\newpage
%------------------------------------------------------------------------------
\subsection{6D Het/F-duality}
%------------------------------------------------------------------------------
This section contains a review of some aspects\footnote{We will not 
focus on nonperturbative 
effects in 6d, such as, strong coupling singularities and tensionless strings.}
of 6D het/F-duality, however,
it is devoted to providing us with the necessary 'six-dimensional'
tools to be used in section 6.3 and 6.4 where 
we will construct four-dimensional
dual heterotic/F-theory pairs with $N=1$ supersymmetry, by a ${\bf Z}_2$
modding of $N=2$ models which are obtained from 6d ones by $T^2$ 
compactification.
%------------------------------------------------------------------------------
\subsubsection{Heterotic string on $K3$}
%------------------------------------------------------------------------------
We will compactify the heterotic string from ten to six-dimensions on 
($K3$,$V$) with $V=SU(n)$ vector bundle, which leads to $N=1$ supergravity
in six-dimensions. We start with the heterotic string on $T^2$ in 
eight-dimensions and vary over an additional ${\bf P^1}$ so that the family
of $T^2$'s fits into an elliptically fibered $K3$, i.e. the elliptic
fibers degenerate over 24 points in the base ${\bf P^1}$. The singularity
type of the elliptic fibre should not be worth than I$_1$ in the Kodaira
classification, in order to obtain a smooth Weierstrass model. We can 
describe the elliptic $K3$ by its Weierstrass equation 
\beqa
y^2=x^3+g_2(z)x+g_3(z)
\eeqa
where $g_2$ and $g_3$ are the polynomials of degree 8 resp. 
12 on the base.

The six-dimensional massless spectrum contains the supergravity multiplet
$G_{6}$ with a graviton $g_{\mu\nu}$, a Weyl gravitino $\psi_{\mu}^{+}$ and 
a self-dual anti-symmetric tensor $B_{\mu\nu}^{+}$; the hypermultiplet $H$
includes a Weyl fermion $\chi^{-}$ and four real massless 
scalar fields $\varphi$;
the tensor multiplet $T$ with an self-dual antisymmetric tensor 
$B_{\mu\nu}^{-}$, a real massless scalar field ${\phi}$ and a 
Weyl spinor $\psi^{-}$; 
finally,
the Yang-Mills multiplet $V$ contains a vector $A_{\mu}$ and a gaugino 
$\lambda^{+}$.\\ \\ 
${\bullet}$ $G_{6}$$=(g_{\mu\nu}, \psi_{\mu}^{+},B_{\mu\nu}^{+})$
$:\ \ (1,1)+2(\frac{1}{2},1)+(0,1)$\\ 
${\bullet}$ $H$=$(\chi^{-},4\varphi)$
$:\ \ 2(\frac{1}{2},0)+4(0,0)$\\ 
${\bullet}$ $T$=$(B_{\mu\nu}^{-},\phi,\psi^{-})$
$:\ \ (1,0)+(0,0)+2(\frac{1}{2},0)$\\ 
${\bullet}$ $V$=$(A_{\mu},\lambda^{+})$
$:\ \ (\frac{1}{2},\frac{1}{2})+2(0,\frac{1}{2})$\\ \\
where the massless representations of $N=1$ supersymmetry are labeled by their
$Spin(4)\sim SU(2)\times SU(2)$ representations.

We get moduli fields from hyper and tensor multiplets. Therefore one 
expects the moduli space to be in the form 
\beqa
\M=\M_H\times \M_T
\eeqa
where $\M_H$ is a quaternionic K\"ahler manifold and $\M_T$ is a Riemannian
manifold, their dimensions are given below.  

Since the supergravity is chiral, there are constaints on the allowed spectrum,
due to gauge and gravitational anomaly cancellation conditions. The anomaly
can be characterized by the exact anomaly eight-form $I_8=dI_7$, given by
\cite{AW}
\beqa
I_8=\alpha trR^4 +\beta (trR^2)^2+\gamma trR^2trF^2+\delta(trF^2)^2
\eeqa
where $F$ is the Yang-Mills two-form, $R$ is the curvature two-form and 
$\alpha$, $\beta$, $\gamma$, $\delta$ are real coefficients depending on the
spectrum of the theory. The gravitational anomaly can only be cancelled if one
requires $\alpha=0$, which leads to \cite{GSWest} the spectrum constraint
\beqa
n_H-n_V+29n_T=273
\eeqa 
where $n_H$ and $n_V$ denote the total number hyper multiplets and vector
multiplets respectively. 
In order to employ a Green-Schwarz mechanism to cancel the remaining 
anomaly, the remaining anomaly eight-form has to factorize, $I_8\sim I_4
\wedge{\tilde I}_4$ with 
\beqa
I_4=trR^2-\sum_i \upsilon_i (trF^2)_i, \ \ \ \ \
{\tilde I}_4=trR^2-\sum_i {\tilde \upsilon}_i(trF^2)_i
\eeqa
where $\upsilon_i$, ${\tilde\upsilon}_i$ are constants which depend on the 
gauge group \cite{GSWest},\cite{S},\cite{Erl}.

Further, if we denote the heterotic string coupling by 
$\lambda^2=e^{2\phi}$, then the fact that the anomaly eight-form 
factorizes, implies \cite{S} that the gauge kinetic term contains a term of the
form
\beqa
\L\propto (\upsilon e^{-\phi}+{\tilde\upsilon}e^\phi)trF^2 
\eeqa
which at finite values of the heterotic coupling 
$e^{-2\phi}=\frac{-\tilde\upsilon}{\upsilon}$ becomes zero and therefore leads
to a phase transition.

To cancel the remaining anomaly, the Green-Schwarz mechanism is  
employed by defining a modified field strength $H$ for the antisymmetric 
tensor $H=dB+\omega^L-\sum_i \upsilon_i \omega_i^{YM}$ with $\int_{K3}dH=0$ 
and $dH=I_4$. $\omega^L$ denotes the Lorentz-Chern-Simons form and 
$\omega_i^{YM}$ is the Yang-Mills Chern-Simons form. So, one gets the condition
\beqa
\sum_i n_i=\sum_i\int_{K3}(trF^2)_i=\int_{K3} R^2=24
\eeqa
Turning on the instanton numbers $n_i$ $(i=1,2)$ in both $E_8$'s, with
\beqa
n_1+n_2=24,
\eeqa
breaks the gauge group to some subgroup $G_1\times G_2$ of $E_8\times E_8$.  

In addition, one has the possibility of turning on a number 
$n_5$ of five-branes
\cite{SW03},\cite{DMW} which are located at points on $K3$ where each point 
contributes a hypermultiplet (i.e. the position of each five-brane is 
parametrized by a hypermultiplet). 
The occurrence of 
nonperturbative five-branes in the vacuum then leads to the generalized 
anomaly cancellation condition
\beqa
n_1+n_2+n_5=24.
\eeqa
Further, the occurrence of such five-branes changes the number of tensor
multiplets in the massless spectrum, since on the world sheet 
theory of the five-brane a massless tensor field lives, one has
\beqa
n_T=1+n_5.
\eeqa 
So, the coulomb branch of the six-dimensional theory is parametrized by the 
scalar field vev's of the tensor multiplets.
If $n_5=0$, then we have a purely perturbative heterotic compactification with
one tensor multiplet, where the corresponding scalar field is the heterotic 
dilaton.  

In order to obtain the complete massless spectrum, we have also to take 
into account the number of bundle moduli from compactification
on $K3$, since they lead to additional gauge neutral hypermultiplets. 
In particular the number of bundle moduli
for $V=SU(n)$ is given by (c.f. section 3)
\beqa
h^1(K3,{\rm{End}}(V))=2\int_{K3}c_2(V)n+1-n^2
\eeqa
So, the quaternionic dimension of the instanton moduli space of $k$ instantons
is given by
\beqa
\dim_{Q}(\M_{inst})&=&nc_2(V)-(n^2-1)\nonumber\\
&=&c_2(H)k-\dim H
\eeqa
where $c_2(H)$ is the dual Coxeter number of the commutant $H$ in 
$G\times H\in E_8$. 
Thus the Higgs branch is parametrized by $n_H$ hypermultiplets 
\beqa
n_H=20+n_5+\dim_Q(\M_{inst})
\eeqa
Now, if $E_8$ is broken to the subgoup $G$ by giving gauge fields
on $K3$ an expectation value in $H$, i.e. $G\times H \in E_8$ is a
maximal subgroup, then the part of the spectrum arising from the Yang-Mills
multiplet can be determined as follows:   

The number of vector multiplets is given by the dimension of the
adjoint representation of the gauge group, since the vector multiplets
belong to the adjoint representation, so we have
\beqa
n_V=\dim({\rm{adj}}(G)).       
\eeqa
Further, the hyper multiplets belong to some representation $M_i$ of $G$. Note
that CPT invarince requires that $M_i$ is real. Denoting by $R_i$ the
representations in $H$ in the decomposition $adj(E_8)=\sum_i(M_i,R_i)$.
   
To determine the number of charged 
hypermultiplets, we consider an $H$-bundle $V$ with fibre in an 
irreducible representation $R_i$ of the structure group. Form the 
Dolbeault index on $K3$ on gets \cite{EGHanson}, \cite{GSWest}
\beqa
\chi(T,V)&=&\sum_{i=0}^{2}(-1)^i h^i(T,V)=\int_{K3}td(T)\wedge ch(V)\nonumber\\
         &=&2\dim(R_i)-\int_{K3} c_2(V){\rm{index}}(R_i).
\eeqa
where $T$ denotes the tangent bundle of $K3$.
Since our bundle is stable, we have $h^{0}(T,V)=0$ and by
Serre duality $h^2(T,V)=0$ or, to say it differently, if $V$ admits 
nonzero global sections then the structure group would be a subgroup
of $H$, but we are interested in strict $H$ bundles. So, we get
\footnote{Note that, in the spectral cover construction of
$V$ one has in the case of $K3$ the class of the spectral curve given by
$C=n\sigma-\pi^*\eta=n\sigma-\pi^*\pi_*c_2(V)=n\sigma-c_2(V)E$. The 
intersection number of $C$ with $\sigma$ of 
$Z\rightarrow P^1$ is given by $C\cdot\sigma=n\sigma^2-c_2(V)E\sigma$ 
and recall
that $E\cdot\sigma=1$ and $\sigma^2=-\sigma c_1(\L)=-2$, we find that
$C\cdot\sigma=c_2(V)-2n$. We thus learn that $h^1(V)$ gets a contribution 
whenever $C$ intersects $\sigma$.}  
\beqa
-h^1(T,V)=2\dim(R_i)-\int_{K3} c_2(V){\rm{index}}(R_i) 
\eeqa
which leads to a condition (c.f.\cite{asp137}) for any irreducible 
representation $R_i$
\beqa
c_2(V)\geq\frac{2\dim(R_i)}{\rm{index}(R_i)}.
\eeqa
Thus, for the number of hypermultiplets in the representation $M_i$, we get
(count quaternionic)
\beqa
N_{M_i}=\frac{1}{2}\int_{K3}c_2(V){\rm{index}}(R_i)-\dim(R_i)
\eeqa\\ \\
{\bf{Examples:}} We start with $E_8\times E_8$ heterotic 
string on $K3$ with $SU(2)$ bundles 
with instanton numbers $(n_1,n_2)=(12-n,12+n)$ embedded into the $E_8$'s 
respectively, where $n_1+n_2=24$. If $0\leq n\leq 8$ then the resulting gauge 
group is $E_7\times E_7$ and with the above index one computes the number
$N_{\bf{56}}$ of massless hypermultiplets  
\beqa
\quad
\frac{1}{2}(8-n)({\bf{56,1}})+\frac{1}{2}(8+n)({\bf{1,56}}).
\quad
\eeqa
In addition one has a number of gauge neutral hypermuliplets coming from
the bundle moduli 
\beqa
\dim_Q(\M_{inst})&=&\dim_Q(\M_{inst}^{n_1})+
                \dim_Q(\M_{inst}^{n_2})\nonumber\\
                &=&(21+2n)+(21-2n)\nonumber\\
                &=&42
\eeqa
and $h^{1,1}(K3)=20$ universal hypermultiplets of $K3$, thus leading
to $62({\bf{1,1}})$ gauge singlets. 
It was shown \cite{MV1}, \cite{MV2}, \cite{Asp118} that complete Higgsing 
of $E_7\times E_7$ is possible for $n=0,1,2$, in particular, the cases $n=0$
and $n=2$ are equivalent. The case $n=12$ is the standard embedding, the 
tangential bundle embedded into one $E_8$, with $c_2(V)=c_2(K3)=24$, one 
gets $E_8\times E_7$ with the massless hypermultiplets 
\beqa
10({\bf{1,56}})+65({\bf{1,1}})
\eeqa
In this case, we can completely Higgs the $E_7$ getting $10\cdot 56-133=427$
hypermultiplets, leading to 492 gauge neutral hypermultiplets and an unbroken
$E_8$ gauge group, thus we get 
\beqa
n_T&=&1\nonumber\\
n_H&=&492\nonumber\\
n_V&=&248  
\eeqa

Let us turn to the F-theory side!  

\newpage
%------------------------------------------------------------------------------

\subsubsection{F-theory on Calabi-Yau threefold}

%------------------------------------------------------------------------------
Let us consider F-theory on Calabi-Yau threefold. We start with 
F-theory in eight-dimensions on an elliptic $K3\rightarrow {\bf P}^1_f$ 
\cite{V}. Variation of the eight-dimensional data over an 
additional ${\bf P}^1_b$ leads to a family of 
elliptic $K3$ surfaces,
which should then fit into a Calabi-Yau threefold $X_3$.

More precisely, $X_3$ should admit an 
elliptic fibration as well as a $K3$ fibration, in
particular, the generic $K3$ fibre should agree in the large volume limit of 
${\bf P}^1_b$ with the heterotic $K3$ in six-dimensions. Thus $X_3$ must be an
elliptic fibration over a two-dimensional base $B_2$ which itself has a
${\bf P}^1_f$ fibration over ${\bf P}^1_b$, i.e. over the rational ruled 
Hirzebruch surface $F_n$ \cite{MV1}, \cite{MV2}.

In the following, we will denote the elliptic fibered Calabi-Yau threefold
over $F_n$ by $X^3_n$ and denoting by $z_1$ and $z_2$ the complex coordinates 
of ${\bf P}^1_b$ respectively ${\bf P}^1_f$. 

The Weierstrass model for $X^3_n$ can be written as \cite{MV1}, \cite{MV2}:
\beqa
X^3_n:\quad 
y^2=x^3+\sum_{k=-4}^4f_{8-nk}(z_1)z_2^{4-k}x+\sum_{k=-6}^6g_{12-nk}(z_1)
z_2^{6-k}.
\label{curve}
\eeqa
where $f_{8-nk}(z_1)$, $g_{12-nk}(z_1)$ are polynomials of degree
$8-nk$, $12-nk$ respectively, and the polynomials with negative degrees
are identically set to zero. So we see that the Calabi-Yau
threefolds
$X^3_n$ are $K3$ fibrations over ${\bf P}^1_b$ with coordinate $z_1$; the 
$K3$ fibres themselves are elliptic fibrations over the ${\bf P}^1_f$ with
coordinate $z_2$.

The Hodge numbers $h^{(2,1)}(X^3_n)$, 
of $X^3_n$, are then given by the number of parameters
of the curve (\ref{curve}) minus the number of possible reparametrizations.
The Hodge numbers $h^{(1,1)}(X^3_n)$ are determined by the Picard number
$\rho$ of the $K3$-fibre of $X^3_n$ as
\beqa
h^{(1,1)}(X^3_n)=1+\rho.\label{h11x3}
\eeqa
Furthermore, the heterotic string coupling was identified in the 
eight-dimensional duality with the size $k_f$ of the ${\bf P^1_f}$. 
Now, in six-dimensions one has \cite{MV1},\cite{MV2}
\beqa
e^{-2\phi}=\frac{k_b}{k_f}
\eeqa
where $k_b$ denotes the K\"ahler class of the base ${\bf P^1_b}$. In 
particular, it was shown \cite{MV1} that there is a bound for the heterotic 
coupling constant due to the K\"ahler cone of the Hirzebruch 
surface $F_n$, given by
\beqa
\frac{k_b}{k_f}\ge \frac{n}{2}
\eeqa
Since on the heterotic side there occur phase transitions at 
$\frac{-\tilde\upsilon}{\upsilon}$, one was led to identify \cite{MV1}
\beqa
\frac{-\tilde\upsilon}{\upsilon}=\frac{n}{2}
\eeqa
This then lead to the duality conjecture between the heterotic string 
on $K3$ with an instanton distribution $(12+n,12-n)$ between both $E_8$'s
and F-theory compactified on Calabi-Yau threefold with $F_n$ as base \cite{V}.
To test these dualities on has to assume that, if the heterotic string 
has an unbroken gauge group $G$ then on the F-theory side the elliptic 
fibration must have a singularity of type $G$\footnote{Note that $G$ is 
always simply laced (ADE groups). But if there is a monodromy action
on the singularity, which is an automorphism of the root latice, then the 
singularity does not correspond to these simply laced groups but to a quotient 
of them and these groups are non-simply laced. These singularities are called
non-split (in contrast, they are called split, if $G$ has only inner 
automorphisms)\cite{Asp131},\cite{BIK}.}.

Let us now recall \cite{V},\cite{MV1},\cite{MV2} how the Hodge numbers 
of $X^3_n$ determine the spectrum of the $F$-theory compactifications.

The number of tensor multiplets $n_T$ is given by the 
number of K\"ahler deformations of the two dimensional type IIB base $B_2$
except for the overall volume of $B_2$, one has
\beqa
n_T=h^{11}(B_2)-1.
\eeqa
Since tensor fields become abelian $N=2$ vector fields upon further
compactification to four-dimensions on $T^2$ and the four-dimensional F-theory
becomes equivalent to type IIA on $X_3$, one is lead to the number of 
four-dimensional abelian vector fields in the Coulomb phase 
$n_T+r(V)+2=h^{11}(X_3)$, where $r(V)$ denotes the rank of the 
six-dimensional gauge group and the 2 arises from the $T^2$ compactification. 
This then leads to: 
\beqa
r(V)=h^{11}(X_3)-h^{11}(B_2)-1. 
\eeqa 
The number of hyper multiplets, which are neutral under the abelian gauge
group, is given in four as well as in six-dimensions by: 
\beqa
n_H=h^{21}(X_3)+1
\eeqa
where the 1 is coming from the freedom in varying the overall volume of $B_2$.
These numbers are constrained by the anomaly cancellation condition 
\beqa
n_H-n_V+29n_T=273
\eeqa 
which provides a check in the spectrum of matter in F-theory.\\ \\
{\bf{Example 1:}} On the heterotic side we had $21+2n$ instantons
in one $E_8$ (resp. $21-2n$ in the other $E_8$). To obtain these numbers 
from F-theory we concentrate at gauge groups, appearing at $z_1=0$. From
the Kodaira classification we find, that the polynomials $f$ and $g$ have to 
vanish on the base to orders $3$ and $5$. One finds \cite{MV1} 
the equation $y^2=x^3+x(z^4f_8+z^3f_{8+n})+(...+z^6g_12+z^5g_{12+n})$ and 
counting complex deformations preserving the singular locus leads to
$(13+n)+(9+n)-1=21+2n$ parameters in agreement with the number of neutral
hypermultiplets (Note that the $-1$ comes from the rescaling of the 
$z_1$ coordinate). Further, we can read of charged matter from the 
discriminant near the $E_7$ locus $z_1=0$, which is 
given by $\Delta=z_1^9(4f^3_{8+n}(z_2)+o(z_1))$. Thus 
we see that $f$ vanishes on $z_1=0$ at $8+n$ points,
corresponding to $8+n$ 7-branes intersecting the one corresponding to $E_7$
\footnote{Note that the types of singularities over these extra branes 
do not necessarily correspond to an extra gauge group enhancement.}. So one 
expects that each charged $\frac{1}{2}$-hypermultiplet in the ${\bf{56}}$
is localized at the points of collision of the divisors\footnote{Katz and 
Vafa \cite{KV} derived, purely in F-theory, the chared matter content 
without making any use of duality with the heterotic string.}.     
\\ \\
{\bf{Example 2:}} The standard embedding on the heterotic side 
($n=12$) leads to $E_8\times E_7$
and complete higgsing of $E_7$ leads to an unbroken $E_8$. To obtain $E_8$ 
gauge symmetry from F-theory on $X_3\rightarrow F_{12}$, we need a 
section $\theta: F_{12}\rightarrow X_3$
of $E_8$ singularities along a codimension one locus in $F_{12}$, 
i.e. we need type II$^*$ singular fibers
along a curve in the Hirzebruch surface $F_{12}$. 
In $F_n$ we have two 'natural'
curves given by the zero section $S_0$ and the section at infinity 
$S_{\infty}=S_0+nf$ of the ${\bf P}^1$ bundle. Let us localize the II$^{*}$
fibers along $S_0$ with $S^2_0=-n$ (also one has $S_0\cdot f=1$). Now, we 
can decompose the discriminant 
$\Delta$ into two components: $\Delta=\Delta_1+\Delta_2$, where $\Delta_1$ 
denotes the component with I$_1$ fibers and $\Delta_2$ has II$^*$ fibres. 
Each component is characterized by the order of vanishing of some polynomials 
as $\Delta$ itself. Recall that $\Delta$ has the class 
$\Delta=-12K_B$, resp. $F=-4K_B$ and $G=-6K_B$. 
Similarly we can denote the class of $\Delta_2$ by $\Delta_2=10S_0$, resp. 
$F_2=4S_0$ and $G_2=5S_0$. Thus, $\Delta_1$ has the class 
$\Delta_1=\Delta-\Delta_2$, resp. $F_1=F-F_2$ and 
$G_1=G-G_2$. With the canonical bundle of the Hirzebruch surface $F_n$,
$K_{F_n}=-2S_0-(2+n)f$ we get (set $n=12$): 
$\Delta_1=14S_0+168f$, resp. $F_1=4S_0+56f$ and $G_1=7S_0+84f$, so describing
the locus of I$_1$ singularities. Further, we have $\Delta_1\cdot \Delta_2=0$
which means that the two components do not intersect, otherwise we must 
blow-up within the base in order to produce a Calabi-Yau threefold 
\cite{AWAL108}. Since we have no additional blow-up's in the base, we get
$h^{1,1}(B_2)=2$ (one from the base ${\bf{P}}^1$ and one from the fiber 
${\bf{P}}^1$ in $F_n$), thus we get $n_T=1$. The number of K\"ahler 
deformations of $X_3$ are given by $h^{1,2}(X_3)=2+8+1=11$ (2 from the 
base, 8 from the resolution of the II$^*$ singular fibers (c.f.\cite{Barth})).
The number of complex structure deformations is now equals 
$h^{2,1}(X_3)=h^{1,1}(X_3)-\frac{1}{2}\chi(X_3)$. Since only the singular
fibers contribute to $\chi(X_3)$, we get 
$\chi(X_3)=\chi(\Delta_1)\chi({\rm{I}}_1)+\chi(S_0)\chi({\rm{II}}^*)$.
Now $\Delta_1$ is a curve in the base, which has cusp singularities at 
$F_1\cdot G_1=392$ points, so 
applying the standard Pl\"ucker formula, we find 
$-\Delta_1(\Delta_1+K_{F_{12}})+2(392)= -1372$.
To determine $\chi(\Delta_1)$ we have to be careful with the cusp contribution,
since over any cusp the elliptic fiber is of type II. Thus each cusp 
contributes with $\chi(II)\chi(F_1\cdot G_1)$ to $\chi(X_3)$. Therefore 
one gets, the corrected formula  
\beqa
\chi(X_3)&=&\chi({\rm{I}}_1)(\chi(\Delta_1)-\chi(F_1\cdot G_1))+\chi({\rm{II}})
\chi(F_1\cdot G_1)+\chi(S_0)\chi({\rm{II}}^*)\nonumber\\
&=&-960
\eeqa 
using the fact that $\chi(S_0)\chi({\rm{II}}^*)=2\cdot 10$, where the 
2 is the Euler characteristic of $S_0={\bf{P}}^1$ of $F_n$ and the 
$\chi({\rm{II}}^*)=2\cdot 9-8=10$ (from the resolution tree of II$^*$ 
singularity, one has 9 ${\bf{P}}^1$'s intersecting in 8 points 
(c.f.\cite{Barth})). Thus we get 
the number of complex deformations $h^{2,1}=491$ and so $n_H=492$ 
hypermultiplets. So we end up with 
the F-theory spectrum
\beqa
n_T&=&1\nonumber\\
n_H&=&492\nonumber\\
n_V&=&248.  
\eeqa

\newpage
%------------------------------------------------------------------------------
\subsection{4D Het/F-duality}
%------------------------------------------------------------------------------

The four-dimensional heterotic/F-theory duality picture can be 
established by considering $Z$ as an elliptic fibration $\pi:Z\rightarrow B_2$
with a section $\sigma$, where $B_2$ is a twofold base; $Z$ can be represented
as a smooth Weierstrass model. $X$ was considered as being elliptically
fibered over a threefold base $B_3$, which is rationally ruled, i.e. there
exists a fibration $B_3\rightarrow B_2$ with ${\bf{P}}^1$ fibers 
because one has to
assume the fourfold to be a $K3$ fibration over the twofold base $B_2$ in order
to extend adiabatically the 8D duality between the heterotic string on $T^2$
and F-theory on $K3$ over the base $B_2$.
\\
During invastigating F-theory compactifications with $N=1$
supersymmetry in four dimensions on a Calabi-Yau fourfold $X$
there was established the necessity
of turning on a number $n_3=\chi (X)/24$ of spacetime-filling threebranes 
for tadpole cancellation \cite{SVW}. This should be compared with a
potentially dual heterotic compactification on an elliptic Calabi-Yau $Z$
with vector bundle $V$ embedded in $E_8\times E_8$.
There the threebranes should correspond to a number $n_5$ of 
fivebranes wrapping the elliptic
fiber. Their necessity for a consistent heterotic compactification
(independent of any duality considerations) was established in the 
exhaustive study done by Friedman, Morgan and Witten on vector budles and
$F$-theory \cite{FMW}. There it was also shown that in the case of an
$E_8\times E_8$ vector bundle $V$, leaving no unbroken gauge group and
corresponding to a smooth Weierstrass model for the fourfold, it is possible
to express $n_3$ and $n_5$ in comparable and indeed matching data on the 
common base $B$. 
\\
In this section we show that in dual $N=1$ string vacua provided by 
the heterotic string on an elliptic Calabi-Yau together with a 
$SU(n_1)\times SU(n_2)$ vector bundle respectively
F-theory on Calabi-Yau fourfold the number of heterotic fivebranes
matches the number of F-theory
threebranes. This extends to the general 
case the work of Friedman, Morgan and Witten, who treated the case of 
$E_8\times E_8$ bundle.
\\
Furthermore, it will be presented a complete matching of the number of 
gauge neutral chiral multiplets obtained in heterotic and F-theory. 
%------------------------------------------------------------------------------
\subsubsection{Heterotic string on Calabi-Yau threefold}
%------------------------------------------------------------------------------

{\bf{The threefold $Z$:}} Let us compactify the heterotic string on a smooth
elliptic Calabi-Yau threefold $Z\rightarrow B_2$. We assume that the elliptic 
fibration has only one section so that $h^{1,1}(Z)=h^{1,1}(B_2)+1=c_2(B_2)-1$.
Recall from {\it section 3}, the Euler characteristic of $Z$ is given by
\beqa
\chi(Z)=-60\int_{B_2}c_1^2(B_2)=2(h^{1,1}(Z)-h^{2,1}(Z))
\eeqa
Using Noethers formula $1=\frac{c_1^2(B_2)+c_2(B_2)}{12}$, we can write the
number of complex and K\"ahler deformations as
\beqa
h^{1,1}(Z)=11-c_1^2(B_2), \ \ \ \
h^{2,1}(Z)=11+29c_1^2(B_2).
\eeqa
{\bf{The bundle $V$:}} In addition to $Z$ we have to specify a vector 
bundle $V$ over $Z$. We are interested in $V=E_8\times E_8$ and 
$SU(n_1)\times SU(n_2)$ in order to break 
the gauge group completely, respectively, to some subgroup of $E_8\times E_8$. 
In what follows, we consider vector bundles $V$ which are $\tau$-invariant, 
i.e. with no additional 'twists' ($\pi_*(\gamma^2)=0$).
Let us collect the necessary bundle data for $V=V_1\times V_2$:\\ \\
\underline{$E_8\times E_8$:} We find for the number of bundle moduli
\beqa
h^1(Z,{\rm{ad}}(V))=16-4((\lambda(V_1)+\lambda(V_2))-(\eta_1+\eta_2)
\sigma)+(\eta_1+\eta_2)c_1+2n_0 
\eeqa
and the fundamental characteristic class is given by
\beqa
\lambda(V_1)+\lambda(V_2)=(\eta_1+\eta_2)\sigma-15(\eta_1^2+\eta_2^2)+
135(\eta_1+\eta_2) c_1-620c_1^2.
\eeqa\\ 
\underline{$SU(n_1)\times SU(n_2)$:} We find for the number of bundle moduli
\beqa
h^1(Z,{\rm{ad}}(V))=rk-4((c_2(V_1)+c_2(V_2))-(\eta_1+\eta_2)
\sigma)+(\eta_1+\eta_2)c_1+2n_0 
\eeqa
with $rk=n_1+n_2-2$ and the second Chern class 
\beqa
c_2(V_1)+c_2(V_2)=(\eta_1+\eta_2)\sigma+(\omega_1+\omega_2)
\eeqa
{\bf{The anomaly:}} We already mentioned in {\it section 2.2} that the 
perturbative anomaly cancellation condition $c_2(V)=c_2(Z)$ will be modified,
due to the occurrence of nonperturbative fivebranes in the vacuum. We have 
\beqa
E_8 &:&   \lambda(V_1)+\lambda(V_2)+[W]=c_2(TZ)\\
SU(n)&:&  c_2(V_1)+c_2(V_2)+[W]=c_2(TZ)
\eeqa
where $[W]$ denotes the cohomology class of the five-branes. 
\\
{\bf{The moduli:}} Now, we can determine the number of moduli, 
which lead to a number $C_{het}$ of $N=1$ neutral chiral (resp.
anti-chiral) multiplets 
\beqa
C_{het}=h^{2,1}(Z)+h^{1,1}(Z)+h^1(Z,{\rm{ad}}(V)).
\eeqa
{\bf Remark:} In general, heterotic string compactifications 
involve a number of $N=1$ chiral matter 
multiplets, which are charged under the unbroken gauge group. This number
is given by (c.f. {\it section 3.2}), using 
$N_{gen}=\frac{1}{2}\mid{\int_{Z} c_3(V)}\mid$ we get  
\beqa
C^c_{het}=N_{gen}+2{\bar n}_{gen}.
\eeqa
Now let us turn to the F-theory side!
\\ 
%------------------------------------------------------------------------------
\subsubsection{F-Theory On Calabi-Yau fourfold}
%------------------------------------------------------------------------------
Let us compute for a {\it general} Calabi-Yau fourfold $X_4$ 
the spectrum of massless $N=1$ superfields in the F-theory compactification,
from the Hodge numbers of $X_4$.
Just as in the six-dimensional case the Hodge numbers of the
type IIB bases $B_3$, i.e. the details of the elliptic fibrations, 
will enter the numbers of massless fields.
So consider first the compactification of the type IIB string from ten
to four dimensions on the spaces $B_3$. Abelian $U(1)$ $N=1$ vector multiplets
arise from the dimensional reduction of the four-form antisymmetric
Ramond-Ramond tensor field $A_{MNPQ}$ in ten dimensions; therefore we expect
that the rank of the four-dimensional gauge group, $r(V)$, gets contributions
from the $(2,1)$-forms of $B_3$ such that $h^{(2,1)}(B_3)$ contributes
to $r(V)$. Chiral (respectively anti-chiral) $N=1$ multiplets,
which are uncharged under the gauge group, arise from from
$A_{MNPQ}$ with two internal Lorentz indices as well from the
two two-form fields $A^{1,2}_{MN}$ (Ramond-Ramond and NS-NS) with zero or
two internal Lorentz indices. Therefore we expect that the number of
singlet chiral fields, $C$, receives contributions from $h^{(1,1)}(B_3)$.
On the other hand we can study the F-theory spectrum in three dimensions
upon futher compactification on a circle $S^1$. This is equivalent to
the compactification of 11-dimensional supergravity on the same $X_4$.
(Equivalently we could also consider the compactifications of the
IIA superstring on $X_4$ to two dimensions.) So in three dimensions the 
11-dimensional three-form field $A_{MNP}$ contributes
$h^{(1,1)}(X_4)$ to $r(V)$ and $h^{(2,1)}$ to $C$. In addition,
the complex structure deformations of the 11-dimensional metric contributes 
$h^{(3,1)}(X_4)$ chiral fields. (These fields arise in analogy to the
$h^{(2,1)}$ complex
scalars which describe the complex structure of the metric when compactifying
on a Calabi-Yau threefold from ten to four dimensions.)
Since, however, vector and chiral fields are
equivalent in three dimensions by Poincare duality, this implies that the
sum $r(V)+C$ must be independent from the Hodge numbers of the type IIB bases 
$B^3$. Therefore, just in analogy to the
six-dimensional F-theory compactifications, the following formulas for
the spectrum of the four-dimensional F-theory models on Calabi-Yau
fourfolds are expected [\cite{CL},\cite{mohri},\cite{ACL}]: 
namely for the rank of
the $N=1$ gauge group we derive 
\beqa
r(V)=h^{(1,1)}(X_4)-h^{(1,1)}(B_3)-1+h^{(2,1)}(B_3) 
\eeqa
and for the number $C_F$ of $N=1$ neutral chiral 
(resp. anti-chiral) multiplets we get
\beqa
C_F&=&h^{(1,1)}(B_3)-1+h^{(2,1)}(X_4)-h^{(2,1)}(B_3)+h^{(3,1)}(X_4)
\nonumber\\
 &=&h^{(1,1)}(X_4)-2+h^{(2,1)}(X_4)+h^{(3,1)}(X_4)-r(V)\nonumber\\
 &=&
 \frac{\chi}{6}-10+2h^{(2,1)}(X^4)-r(V).
\eeqa
Note that in this formula we did not count the chiral field which 
corresponds to the dual heterotic dilaton.
Further to get a consistent F-theory compactification on $X_4$ we have to 
include a number $n_3$ of space-time filling threebranes in the vacuum 
\cite{SVW}.
\beqa
n_3=\frac{\chi(X_4)}{24}
\eeqa\\ 
{\bf{Note:}} We do not turn on gauge bundles inside the sevenbrane, moreover,
we assume that no fourflux is turned on(since the fourflux in F-theory is 
related to the 'twists' of the line bundle on the spectral cover: 
$\pi_*(\gamma^2)=-G^2$ \cite{DC}.)

%------------------------------------------------------------------------------
\subsubsection{The duality}
%------------------------------------------------------------------------------
To establish the four-dimensional het/F-theory duality we will implement 
the adiabatically extended duality by the following specification \cite{FMW}:
as $X_4$ is assumed
to be a $K3$ fibration over $B_2$ it follows that $B_3$, the threefold base 
of the F-theory elliptic fibration, is a ${\bf{P}}^1$ fibration over $B_2$; 
this 
fibration structure is described by assuming the ${\bf{P}}^1$ bundle over $B_2$
to be a projectivization of a vector bundle $Y={\cal O}\oplus {\cal T}$, with
${\cal T}$ a line bundle over $B_2$ (c.f. {\it section 3}); then 
the cohomology class $t=c_1({\cal T})$
encodes the ${\bf{P}}^1$ fibration structure. Now the duality is implemented by
choosing for our bundle
\beqa
\eta_1=6c_1(B_2)+t, \ \ \  \eta_2=6c_1(B_2)-t
\eeqa
where $\eta=\pi_*(\lambda(V))$ in the case of $V=E_8$ and 
$\eta=\pi_*(c_2(V))$ for $V=SU(n)$. 
Note that this specification is the analogue of the relation in the 6D het/F-
theory duality \cite{V} with ($12+n,12-n$) instantons embedded in each $E_8$ 
and F-theory on elliptic fibered threefold over $F_n$.\\ \\  
{\bf{Brane Match:}}\\ 
Let us first recall to what was done by Friedman, Morgan and Witten. They 
considered $E_8\times E_8$ bundle and a smooth Calabi-Yau fourfold. 
With $\eta_{1,2}=6c_1(B_2)\pm t$ one can express the 
fundamental characteristic class of the $E_8$ bundle as
\beqa
\lambda(V_1)+\lambda(V_2)=-80c_1^2+12\sigma c_1-30t^2
\eeqa
and using further $c_2(Z)=c_2+11c_1^2+12\sigma c_1$ one finds the number
of fivebranes \cite{FMW}
\beqa
n_5=c_2+91c_1^2+30t^2
\eeqa
On the F-theory side, one has to express $\chi(X_4)$ in 
terms of base $B_2$ data and finds for the number of threebranes
\beqa
n_3=\int_{B_2}c_2+91c_1^2+30t^2.
\eeqa 
which matches the heterotic fivebranes. 
\\ \\
Now let us see how this matching 
proceeds in the case of $V=SU(n_1)\times SU(n_2)$. 
To achieve this we will adopt a somewhat different technical procedure 
\cite{AC}.
In the case of not having a smooth Weierstrass model there occured already
in \cite{SVW} the difficulty to reduce the fourfold expression
$\chi (X_4)/24$ for $n_3$ to an expression involving only suitable
data of the base $B_3$ of the elliptic F-theory fibration of $X_4$ (not to
be confused with the twofold base $B_2$ of the $K3$ fibration of $X_4$, here
denoted simply by $B_2$, which is visible also on the heterotic side),
which was \cite{FMW} only an intermediate step to reduce the expression
to one involving only twofold base data.
For this reason we express here $\chi (X)$ directly in the Hodge numbers 
of $X_4$ and match then these with the data of the dual heterotic model;
here essential use is made of the index-formula computation.
\footnote{Note that in 
case some of the threebranes have dissolved into finite-sized
instantons in the world-volume gauge theory of the F-theory seven-brane,
this is accompanied by a corresponding five-brane transition on the heterotic
side \cite{BJPS}, \cite{FMW}, \cite{S}.}

Let us start and first compute the number of heterotic five-branes 
$n_5$ using the nonperturbative anomaly cancellation condition 
$n_5=c_2(Z)-(c_2(V_1)+c_2(V_2))$. Here we
have to recall that this formula was derived \cite{FMW}
under the assumption that Z admits a
smooth Weierstrass model, an assumption we adopt here as already stated 
(note that in contrast to the F-theory side, where this
bears essential physical content, it represents on the heterotic side
just a technical assumption).
For explicit evaluation and to establish some notation we give here the 
actual number of fivebranes for $V$ a $SU(n_1)\times SU(n_2)$ bundle.
The second Chern class for a $SU(n)$ bundle
was given by (with $\eta= \pi_*(c_2(V))$ and ${\cal L}$ being some line
bundle over $B_2$)
\beqa
c_2(V)&=&\eta\sigma +\omega\nonumber\\
&=& \eta\sigma -\frac{1}{24}c_1({\cal L})^2(n^3-n)-\frac{n}{8}\eta(
                                      \eta-nc_1({\cal L}))
\eeqa
Using the above relations
for $\eta_1$, $\eta_2$ and $c_2(V)$ we can derive 
(where also $c_1(B)=c_1({\cal L})$ by the 
Calabi-Yau condition for $Z$)
\beqa
c_2(V_{1/2})=6c_1\sigma \pm t\sigma - \frac{1}{24}c_1^2
            (n_1^3-n_1)- \frac{n_1}{8}[36c_1^2 \pm 12c_1\, t
            +t^2 -6n_1c_1^2\mp n_1\, t\, c_1]
\eeqa  
We find for the number of fivebranes
\beqa
n_5&=& c_2+11c_1^2+\frac{1}{24}c_1^2(n_1^3-n_1+n_2^3-n_2)+
       \frac{(n_1+n_2)}{4}(18c_1^2+\frac{t^2}{2})+\nonumber\\
   & & +\frac{(n_1-n_2)}{4}6c_1\, t-
       \frac{(n_1^2+n_2^2)}{8}
       6c_1^2+\frac{(n_2^2-n_1^2)}{8}\, t\, c_1
\eeqa
We can now express the number of F-theory threebranes
$n_3=\chi (X_4)/24$ in terms of heterotic data. Because of
$\chi (X)/6 -8=h^{1,1}(X_4)+h^{3,1}(X_4)-h^{2,1}(X_4)$ 
we have to use the following informations \footnote{
$r$ denotes the rank of the unbroken non-abelian gauge group (do not confuse
it with the rank of the group of the bundle $V$); we furthermore assume that 
we have no further $U(1)$ factors (coming from sections).}
\beqa
h^{1,1}(X_4)&=& h^{1,1}(Z)+1+r=12-c_1^2+r\label{11}\\
h^{2,1}(X_4)&=& n_o\\
h^{3,1}(X_4)&=& h^{2,1}(Z)+I+n_o+1=12+29 c_1^2+I+h^{2,1}(X)\label{22}
\eeqa
where in the first line
we used Noethers formula and
$\chi (Z)=-60\int _{B_2} c_1^2(B_2)$; for the $n_o$ in the second line
see below. Let us now see how the 
expression for $h^{3,1}(X_4)$ emerges. Note that
the moduli space ${\cal M}$ for bundles on $Z$ of dimension $m_{bun}$ 
has a fibration ${\cal M}\rightarrow {\cal Y}$ which corresponds on the 
F-theory side to a fibration of the abelian variety 
$H^3(X_4,{\bf R})/H^3(X_4,{\bf Z})$ of complex dimension $h^{2,1}(X_4)$ over
a part of the space of complex deformations of $X_4$ (cf. \cite{FMW}); the 
remaining complex deformations account for the complex deformations of 
the heterotic Calabi-Yau (+1), i.e.(see below)
\beqa
h^{3,1}(X_4)+h^{2,1}(X_4)=h^{2,1}(Z)+m_{bun}+1\nonumber
\eeqa
Now concerning the contribution of the bundle moduli recall $I$ 
is given by $I=-\sum_{i=o}^3 (-1)^i h^i(Z,{\rm{ad}}(V))_e$
where the subscript "e" (resp. "o") indicates the even (resp. odd) part. As
we have an unbroken gauge group $H$, which is the commutator of the group
$G$ of $V$, one finds $I=n_e-n_o$ corrected by $h^0_e-h^0_o$
denoting by $n_{e/o}$  the number $h^1(Z,{\rm{ad}}(V))_{e/o}$ of 
massless even/odd
chiral superfields and by $h^0_{e/o}$ the number of unbroken gauge group
generators even/odd under $\tau$. The unbroken gauge group is in this $n_3$-
calculation accounted for by the rank contribution in $h^{1,1}(X_4)$ for the 
resolved fourfold.

So one has for the number of bundle moduli $m_{bun}=h^1(Z,{\rm{ad}}(V))=
n_e+n_o=I+2n_o$ and so
gets the announced expression for $h^{3,1}(X_4)$. Furthermore on the 
F-theory side the modes odd under the involution $\tau ^{\prime}$ 
corresponding to the heterotic involution $\tau$ correspond to the 
$h^{2,1}(X_4)$ classes \cite{FMW} (we assume no four-flux was turned on).
Using the formula for the index of $SU(n_1)\times SU(n_2)$ bundle and 
the expressions for $\eta_1$ and $\eta_2$ and $c_2(Z)$
and denoting $rk=n_1+n_2-2=16-r)$ we find for $I$
\beqa
I&=&rk-4(c_2(V_1)+c_2(V_2))+48c_1 \sigma +12c_1^2\nonumber\\
 &=& rk-4(c_2+11c_1^2+12\sigma c_1)+4n_5+
     48c_1 \sigma +12c_1^2\nonumber\\
 &=&rk-(48+28c_1^2)+4n_5\label{5}
\eeqa
One finally gets then with $r=16-rk$ and our expression for $I$ that
\beqa
\chi (X_4)/24&=&2+\frac{1}{4}(12-c_1^2+16-rk+12+29 c_1^2+I)\nonumber\\
&=&2+\frac{1}{4}(40+28c_1^2+I-rk)\nonumber\\
&=&n_5
\eeqa
matching the heterotic value. 

After showing that the number of heterotic fivebranes matches the number
of F-theory threebranes for $E_8$ and $SU(n)$ vector bundle on $Z$, we
now turn to the moduli match.
\\ \\
{\bf{Moduli Match:}}\\
Let us now compare the number of moduli in heterotic and 
F-theory which lead to $N=1$ neutral chiral multiplets. 
Recall, this number was given by  
\beqa
C_{het}&=&h^{2,1}(Z)+h^{1,1}(Z)+h^1(Z,{\rm{ad}}(V))\\
&=&22+28c_1^2+I+2n_0
\eeqa
using $h^{1,1}(Z)=11-c_1^2(B_2)$ resp. $h^{2,1}(Z)=11+29c_1^2(B_2)$. 

On the F-theory side we have 
\beqa
C_F=\frac{\chi(X_4)}{6}-10+2h^{(2,1)}(X^4)-r(V).  
\eeqa
using $\chi(X_4)=6(8+h^{1,1}(X_4)+h^{2,1}(X_4)+h^{3,1}(X_4)-h^{2,1}(X_4))$ and
(\ref{11}-\ref{22}) for the hodge numbers, we get
\beqa
C_F=22+28c_1^2+I+2h^{2,1}(X_4)  
\eeqa
So, if we identify the modes which are odd under the $\tau'$ involution on the
F-theory side with those odd under $\tau$-involution on the heterotic side, 
i.e.
\beqa
n_0=h^{2,1}(X_4).
\eeqa
we get complete matching.

Note that this is possible as long as we assume that no four-flux is turned on
which otherwise would imply that we have to take into account the 
twistings appearing in the spectral cover construction of $V$. The twistings
lead to a multi-component structure of the bundle moduli space 
(c.f.\cite{DC},\cite{BJPS})!

Let us consider an example! The case $V=E_8\times E_8$ on $Z$ which
means on the F-theory side that there is no unbroken gauge group!

The number of geometrical and bundle moduli which leads to the number
$C_{het}$ of $N=1$ neutral chiral multiplets can be written as
\beqa
C_{het}&=&h^{2,1}(Z)+h^{1,1}(Z)+h^1(Z,{\rm{ad}}(V))\nonumber\\
&=&38+360\int_{B_2}c_1^2(B_2)+120\int_{B_2}t^2+2n_0
\eeqa
using $h^{2,1}+h^{1,1}=22+28\int_{B_2}c_1^2(B_2)$ and the number of bundle 
moduli
$h^1(Z,{\rm{ad}}(V))=I+2n_0$ which we express 
in terms of $\eta_{1,2}=6c_1(B_2)\pm t$, one has (c.f.\cite{ACL})
\beqa
I=16+332\int_{B_2}c_1^2+120\int_{B_2}t^2.
\eeqa
The expected number $C_F$ of F-theory neutral chiral multiplets is given by
\beqa
C_F=\frac{\chi(X_4)}{6}-10+2h^{(2,1)}(X_4)-r(V).  
\eeqa
Since we have no unbroken gauge group we have $r(V)=0$ and if we use 
\beqa
\frac{\chi(X_4)}{24}=12+90\int_{B_2}c_1^2(B_2)+30\int_{B_2}t^2
\eeqa
we will find
\beqa
C_F=38+360\int_{B_2}c_1^2(B_2)+120\int_{B_2}t^2+2h^{2,1}(X_4)
\eeqa
and thus getting complete matching (identifying $n_0=h^{2,1}(X_4)$)
\beqa
C_{het}=C_F
\eeqa

\newpage
%------------------------------------------------------------------------------
\section{4D Het/F-Models}
\resetcounter
%------------------------------------------------------------------------------
In the following, we consider $N=1$ dual F-theory/heterotic
string pairs with $B_2$ being the socalled del Pezzo surfaces
$dP_k$, the ${\bf{P}}^2$'s blown up in $k$ points. Then, in our class
of models the three-dimensional bases, $B^3_{n,k}$, are characterized
by two parameters $k$ and $n$, where $n$ encodes the fibration
structure $B^3_{n,k}\rightarrow dP_k$.
Specifically we will discuss the cases $n=0$, $k=0,\dots ,6$, i.e.
$B^3_{0,k}={\bf{P}}^1\times dP_k$, and also the cases $n=0,2,4,6,12$, $k=9$, 
where the 9 points are the intersection of two cubics and 
the $B_{n,9}$ are a fibre product
$dP_9\times_{\bf{P}^1} F_{n/2}$ with common base ${\bf{P}}^1$ 
and $F_{n/2}$ the 
rational ruled Hirzebruch surfaces.
We will show that the massless spectra match on both sides.

The considered class of models is the direct generalization
of the F-theory/heterotic $N=1$ dual string pair with $n=0$, $k=9$ 
(again the 9 points in the mentioned special position),
which were investigated in \cite{CL}, which was a ${\bf Z}_2$ modding
of a $N=2$ model described as F-theory on $T^2\times CY^3$ (also
considered there was the ${\bf Z}_2$ modding of $K3\times K3$). In this model
it was also considered a duality check for the superpotential.

The first chain of models we will discuss here consists then in
heterotic Calabi-Yau's elliptic over $dP_k$ ($k=0,\dots ,3$) with
a {\it general} $E_8\times E_8$ vector bundle resp. F-theory on smooth
Calabi-Yau fourfolds
$X^4_k$ elliptic over $dP_k \times {\bf{P}}^1$. The heterotic
Calabi-Yau of $\chi\neq 0$ varies with $k$ and one can also see
the transition induced from the blow-ups in ${\bf{P}}^2$ among the
fourfolds on the F-theory side (for transitions among $N=1$ vacua cf. also
\cite{KS}.

Then we go on to construct a second chain of four-dimensional
F-theory/heterotic dual string pairs with
$N=1$ supersymmetry by ${\bf{Z}}_2$-modding of corresponding dual pairs 
with $N=2$ supersymmetry.
The resulting Calabi-Yau four-folds $X^4_n$ are $K3$-fibrations over the
del Pezzo surface $dP_9$ (with points in special positions).
On the heterotic side, the dual models are obtained by compactification
on a known Voisin-Borcea Calabi-Yau three-fold with
Hodge numbers $h^{(1,1)}=h^{(2,1)}=19$, where,
similarly as in the underlying $N=2$ models, the 
heterotic gauge bundles over this space are characterized by turning on
$(6+{n\over 2},6-{n\over 2})$ instantons of $E_8\times E_8$.
We work out the Higgsing chains of the 
gauge groups together with their massless
matter content (for example the numbers of 
chiral multiplets in the ${\bf 27}+{\bf\overline{27}}$ representation
of $E_6$) for each model and 
show that the  heterotic spectra
of our models match the
dual F-theory spectra, as computed from the Hodge numbers
of the four-folds $X^4_n$ and of the type IIB base spaces.

Note that in contrast to this second class of models the first chain
(varying $k$, $n=0$) consists of {\it genuine} $N=1$ models.
Moreover, within the first chain the heterotic Calabi-Yau spaces have
non-vanishing Euler numbers, potentially leading to theories
with chiral spectra with respect to  non-Abelian gauge groups which show
up at certain points in the moduli spaces. Hence the transitions in $k$
might connect $N=1$ models with different numbers of chiral fermions.
%------------------------------------------------------------------------------
\subsection{ F-theory on ${X^4_k}\rightarrow 
           {\bf{P}}^1\times {dP}_k$}
%------------------------------------------------------------------------------
Let us start and consider F-theory on a 
smooth elliptically fibered Calabi-Yau fourfold \cite{ACL}
$X^4$ with base $B^{3}_k={\bf P}^1\times {dP}_k$ which can be represented 
by a smooth Weierstrass model over $B^{3}_k$ if the anti-canonical
line bundle $-K_{B}$ over $B^{3}_k$ is very ample 
\cite{grasi}.
The Weierstrass model is can be described as above
by the homogeneous equation $y^{2}z=x^{3}+g_{2}xz^{2}+g_{3}z^{3}$
in a ${\bf P}^2$ bundle $W\rightarrow B^{3}_k$. The ${\bf P}^2$ bundle 
is the projectivization of a vector bundle $K_{B}^{-2}\oplus
K_{B}^{-3}\oplus{\cal O}$ over $B^{3}_k$. 

Most of the 84 known Fano threefolds have a very ample $-K_{B}$ \cite{grasi}. 
18 of them are toric threefolds ${\cal F}_n$ (
$1\le n \le 18$) and are completely classified \cite{its1}, \cite{its2},
\cite{MM}, \cite{MM2}. They were studied in the context of 
Calabi-Yau fourfold compactifications over Fano threefolds \cite{klemm}, 
\cite{mohri}. In particular in \cite{grasi}, it was shown that for 
$k=0,...,6$ over $B^{3}_k$ there exists a smooth Weierstrass model 
(having a section); $B^{3}_k$ with $k=0,1,2,3$ 
correspond to the toric Fano threefolds ${\cal F}_n$ with $n=2, 9, 13, 17$.
The corresponding fourfolds $X^4$ have a $K3$ fibration over ${dP}_k$ 
(cf. \cite{mohri},\cite{mayr}).
First we can determine the Hodge numbers of $B^{3}_k$.
The number of K\"ahler and complex structure parameters of $B^{3}_k$ are 
$h^{11}(B^{3}_k)=2+k$ (where the 2 is comes from the line in ${\bf P}^2$ and 
the class of the ${\bf P}^1)$ and $h^{21}(B^{3}_k)=0$.
For $k=0,...,6$ we can compute the Euler number of X in terms of topological 
data of the base 
\begin{eqnarray}
\chi(X^4_k)=12\int_{B^{3}_k}c_1c_2+360\int_{B^{3}_k}c_{1}^3
\end{eqnarray}
where the $c_{i}$ refer to $B^{3}_k$ (the 360 is related to the
Coxeter number of $E_8$ which is associated with the elliptic fiber
type \cite{klemm}).\\

One finds $\int_{B^{3}_k}c_1c_2=24$ and 
$\int_{B^{3}_k}c_{1}^3= 3c_{1}({\bf P}^1)c_{1}(dP_k)^2= 6(9-k)$ which
leads to 
\begin{eqnarray}
\chi(X^4_k)=288+2160(9-k).
\end{eqnarray}
In the following we restrict to
$k=0,1,2,3$ where one has $h^{21}=0$, \cite{klemm},\cite{mohri}],
$h^{11}(X^4_k)=3+k$, $h^{31}(X^4_k)=\frac{\chi}{6}-8-(3+k)=
28+361(9-k)$ \cite{klemm},\cite{mohri},\cite{SVW}].
Now since we have a smooth Weierstrass model and thus expecting no unbroken 
gauge group, i.e.
\begin{eqnarray}
r(V)&=& h^{11}(X^4_k)-h^{11}(B^3_k)-1+h^{21}(B^3_k)\nonumber\\
    &=& 0
\end{eqnarray}
and for the number of N=1 neutral chiral (resp. anti-chiral)
multiplets $C_{F}$ we get
\begin{eqnarray}
C_{F}&=&h^{11}(B^3_k)-1+h^{21}(X^4_k)-h^{21}(B^3_k)+h^{31}(X^4_k)\nonumber\\
     &=&38+360(9-k)
\end{eqnarray}

\begin{center}
\begin{tabular}{|c||c|c|c|c|c|} \hline
k & $\chi$ & $h^{31}$ & $h^{22}$ & $h^{11}$ & $C_{F}$ 
\\  \hline 
0 & 19728 & 3277 & 13164 &  3  & 3278 \\  \hline
1 & 17568 & 2916 & 11724 &  4  & 2918 \\  \hline
2 & 15408 & 2555 & 10284 &  5  & 2558 \\  \hline 
3 & 13248 & 2194 &  8844 &  6  & 2198 \\  \hline
4 & 11088 & 1833 &  7404 &  7  & 1838 \\  \hline
5 &  8928 & 1472 &  5964 &  8  & 1478 \\  \hline
6 &  6768 & 1111 &  4524 &  9  & 1118 \\  \hline
\end{tabular}
\end{center}

In this table we have made for the cases $k=4,5,6$ the assumption 
$h^{21}(X^4)=0$. With this assumption, 
the matching we will present in the following goes through also in
these cases.

Furthermore, for the number of threebranes $n_3$ we have 

\begin{eqnarray}
n_3=\int_{B^{2}}(c_{2}+91c_{1}^2+30t^2)
=822-90k
\end{eqnarray}
using $t=0$ since we have $B^3_k={\bf P}^1\times {dP}_k$ (recall 
$t$ measures the non-triviality of the ${\bf P}^1$ bundle over 
$B_2={dP}_k$.) 

From the last expression we learn that between each blow up there is a 
threebrane difference of 90. 
Note that Sethi, Vafa and Witten \cite{SVW} had a brane difference of 120
as they blow up in a threefold whereas we do this in $B^2$.
\\
Now let us turn to the heterotic side!
\newpage
%----------------------------------------------------------------------------- 
\subsection{ Heterotic string on ${Z}\rightarrow {dP}_k$}
%-----------------------------------------------------------------------------
To compactify the heterotic string on $Z$,
we have in addition to specify our vector bundle V with fixed
second Chern class. Since we had zero for the rank $r(V)$ of the N=1 
gauge group on the F-theory 
side we have to switch on a $E_{8}\times E_{8}$ bundle that breaks the
gauge group completely.\\ 
Now let $Z$ be a nonsingular elliptically fibered Calabi-Yau threefold over
${dP}_k$ with a section. Recall the Picard group of ${dP}_k$
is $Pic$ ${dP}_k={\bf Z}{\ell}\oplus{\bf Z}{\ell}_i$ where ${\ell}$ denotes 
the class of the line in  ${\bf P}^2$ and ${\ell}_i$, $i=1,...,k$ are the 
classes of the blown up points. The intersection form is 
defined by \cite{Manin}
\begin{eqnarray}
{\ell}^2=1, \ \ \ {\ell}_{i}\cdot{\ell}=0 , 
\ \ \  {\ell}_{i}\cdot {\ell}_{j}=-\delta_{ij}, \ \
\end{eqnarray} 
and the canonical class of ${dP}_k$ is 
\begin{eqnarray}
{\cal K}_{\cal B}=-3{\ell}+\sum_{i=1}^{k}{\ell}_i .
\end{eqnarray}  
Assuming again that we are in the case of a 
general\footnote {i.e. having only 
one section \cite{grasi}, so a typical counterexample would be the
$CY^{19,19}=B_{9}\times _{\bf{P}^1}B_{9}$ with 
the 9 points being the intersection
of two cubics} smooth Weierstrass model one has $h^{11}(Z)=2+k$. So the 
fourth homology of $Z$ is generated by the following divisor
classes: $D_0=\pi^*{\ell}$, $D_i=\pi^*{\ell}_i$ and $S=\sigma({\cal B}_k)$.
The intersection form on $Z$ is then given by

\begin{eqnarray}
S^3=9-k, \ \ D_{0}^2S=1, \ \ D_{i}^2S=-1, \ \ 
                              D_{i}S^2=-1, \ \ D_{0}S^2=-3,  
\end{eqnarray}   

all other triple intersections are equal to zero.\\
The canonical bundle for $Z\rightarrow B$ is given by \cite{MV1},\cite{MV2}

\begin{eqnarray}
{\cal K}_{Z}=\pi^{*}({\cal K}_{B}+\sum a_{i}[\Sigma_{i}])  
\end{eqnarray} 

where $a_{i}$ are determined by the type of singular fiber and $\Sigma_{i}$
is the component of the locus within the base on wich the elliptic curve 
degenerates. In order to get ${\cal K}_{Z}$ trivial, one requires 
${\cal K}_{B}=-\sum a_{i}[\Sigma_{i}]$. \\
Since we are interested in a smooth elliptic fibration, we have to check
that the elliptic fibration does not degenerate worse than with $I_{1}$
singular 
fiber over codimension one in the base which then admits a smooth
Weierstrass model. That this indeed happens in our case of del Pezzo base
can be proved by similar methods as for the $F_{n}$ case in \cite{MV1}.
As $a_{i}=-1$ we have a smooth elliptic fibration.\\
The Euler number of $Z$ is given by (as above) 
$\chi(Z)=-60 \int c_{1}^2(B)$
and\footnote{Note that the transition by 29 in $h^{21}(Z)$ in going from
$k$ to $k+1$ has a well known interpretation if one uses these Calabi-Yau
threefolds as F-theory compactification spaces cf. \cite{CF}}
\begin{eqnarray}
h^{11}(Z)&=&2+k=11-(9-k) \\
h^{21}(Z)&=&(272-29k)=11+29(9-k).
\end{eqnarray}
With the Index of the $E_8\times E_8$ vector bundle
\begin{eqnarray}
I=16+332\int_{B}c_{1}^2(B)+120\int_{B}t^2
\end{eqnarray}  
recalling that in our case the last term vanishes, we find for the number
of bundle moduli (with $n_0=0$ since we have $h^{2,1}(X^4)=0$)
\begin{eqnarray}
I=16+332(9-k)
\end{eqnarray}
and for the number of $N=1$ neutral chiral (resp. antichiral) multiplets
$C_{het}$ we find
\begin{eqnarray}
C_{{\rm het}}&=& h^{21}(Z)+h^{11}(Z)+I\nonumber\\
   &=& 38+360(9-k)
\end{eqnarray}
which agrees with the number of chiral multiplets $C_F$ 
on the F-theory side.
\vspace{1.3cm}
\begin{center}
\begin{tabular}{|c||c|c|c|c|} \hline
k & $\chi$ & $h^{21}$ & $h^{11}$ & $I$\\  \hline 
0 & -540 & 272 &  2  & 3004    \\  \hline
1 & -480 & 243 &  3  & 2672    \\  \hline
2 & -420 & 214 &  4  & 2340    \\  \hline 
3 & -360 & 185 &  5  & 2008    \\  \hline
4 & -300 & 156 &  6  & 1676    \\  \hline
5 & -240 & 127 &  7  & 1344    \\  \hline
6 & -180 &  98 &  8  & 1012    \\  \hline
\end{tabular}
\end{center}
\newpage
%------------------------------------------------------------------------------
\subsection{F-theory on the $K3$-fibred fourfolds $X^4_n$}
%------------------------------------------------------------------------------

%------------------------------------------------------------------------------
\subsubsection{The $N=2$ models: F-theory 
on $X^3_n(\times T^2)$}
%------------------------------------------------------------------------------
We will start constructing the fourfolds $X^4_n$
by first considering F-theory compactified
to six dimensions on an elliptic Calabi-Yau threefold $X^3$, and then
further to four dimensions on a two torus $T^2$, i.e. the
total space is given by $X^3\times T^2$.
This leads to $N=2$ supersymmetry in four dimensions.
As explained in \cite{V}, this four-dimensional F-theory compactification
is equivalent to the type IIA string compactified on the same
Calabi-Yau $X^3$. 
 
To be more specific, let us discuss the cases where the Calabi-Yau
threefolds,
which we call $X^3_n$, are elliptic fibrations over 
the rational ruled Hirzebruch
surfaces $F_n$  as already mentioned in section 5.
The corresponding type IIB base spaces are given
by the Hirzebruch surfaces $F_n$ in six dimensions. $X^3_n$ is given by
(c.f. section 5)
\beqa
X^3_n:\quad 
y^2=x^3+\sum_{k=-4}^4f_{8-nk}(z_1)z_2^{4-k}x+\sum_{k=-6}^6g_{12-nk}(z_1)
z_2^{6-k}.\label{weier}
\eeqa
with $f_{8-nk}(z_1)$, $g_{12-nk}(z_1)$ are polynomials of degree
$8-nk$, $12-nk$ respectively, where the polynomials with negative degrees
are identically set to zero. From this equation we see that the Calabi-Yau
threefolds
$X^3_n$ are $K3$ fibrations over $P^1_{z_1}$ with coordinate $z_1$; the 
$K3$ fibres themselves are elliptic fibrations over the $P^1_{z_2}$ with
coordinate $z_2$.

Recall, $h^{(2,1)}(X^3_n)$, 
counting the number of complex structure
deformations of $X^3_n$, are given by the the number of parameters
of the curve (\ref{weier}) minus the number of possible reparametrizations,
which are given by 7 for $n=0,2$ and by $n+6$ for $n>2$.
On the other hand, the Hodge numbers $h^{(1,1)}(X^3_n)$, which count the number
of  K\"ahler parameters of $X^3_n$, are determined by the Picard number
$\rho$ of the $K3$-fibre of $X^3_n$ as
\beqa
h^{(1,1)}(X^3_n)=1+\rho.\label{h11x3}
\eeqa
Let us list the Hodge numbers of the $X^3_n$ for those cases 
relevant for our following discussion:
\hspace{0.2cm}
\beqa
%\begin{center}
\begin{tabular}{|c||c|c|} \hline
  $n$     & $h^{(1,1)}(X^3_n)$ & $h^{(2,1)}(X^3_n)$ \\ \hline\hline
0  &   3 & 243   \\ \hline
2 &   3 & 243    \\ \hline
4 &    7 & 271 \\ \hline
6 & 9 & 321 \\ \hline
12 & 11 & 491 \\ \hline
\end{tabular}
%\end{center}
\label{tab1}
\eeqa
%\hspace{0.5cm}
Further recall, the Hodge numbers 
of $X^3_n$ determine the spectrum
of the F-theory compactifications. In six dimensions
the number of tensor multiplets $T$ was given 
\beqa
n_T=h^{(1,1)}(B^2)-1.\label{notensor}
\eeqa
The rank of the six-dimensional gauge group is given by
\beqa
r(V)=h^{(1,1)}(X^3_n)-h^{(1,1)}(B^2)-1.\label{novector}
\eeqa
Finally, the number of hypermultiplets $n_H$, which are neutral under the Abelian
gauge group is
\beqa
n_H=h^{(2,1)}(X^3_n)+1.\label{nohyper}
\eeqa
For the cases we are interested in, namely $B^2=F_n$, $h^{(1,1)}(F_n)$ is
universally given by $h^{(1,1)}(F_n)=2$. Therefore one immediately gets
$n_T=1$, which corresponds to the universal dilaton tensor multiplet
in six dimensions, and 
\beqa
r(V)=\rho -2=h^{(1,1)}(X^3_n)-3.\label{novector1}
\eeqa

At special loci in the moduli spaces of the hypermultiplets one
obtains enhanced non-Abelian gauge symmetries. These loci are determined 
by the 
singularities of the curve (\ref{weier}) and were analyzed in detail in
\cite{BIK}. These F-theory
singularities correspond to the perturbative gauge symmetry enhancement
in the dual heterotic models.
%------------------------------------------------------------------------------
\subsubsection{The $N=1$ models: 
F-theory on $X^4_n=(X^3_n\times T^2)/{\bf{Z}}_2$}
%------------------------------------------------------------------------------

Now we will construct  from the $N=2$ $F$-theories on
$X^3_n\times T^2$ the corresponding $N=1$ models on Calabi-Yau four-folds 
$X^4_n$ by a ${\bf{Z}}_2$ modding procedure, i.e.
\beqa
X^4_n={X^3_n\times T^2\over {\bf{Z}}_2}.\label{modding}
\eeqa
%The precise definition of the $Z_2$ modding follows from the corresponding
%operation on the heterotic side (see
%next chapter), as we have already constructed
%the four-fold $X^4_0$ by this procedure
%in \cite{CL}.
First, the ${\bf{Z}}_2$ modding acts as quadratic redefinition
on the coordinate $z_1$, the coordinate
of the base ${\bf{P}}^1_{z_1}$ of the $K3$-fibred space $X^3_n$,
i.e. the operation is $z_1\rightarrow -z_1$. This means that
the modding is induced from the quadratic base map
$z_1\rightarrow \tilde z_1:=z_1^2$ with the two branch points 0 and $\infty$.
So the degrees of the corresponding polynomials $f(z_1)$ and $g(z_1)$
in eq.(\ref{weier}) are reduced by half (i.e. the moddable cases are the
ones where only even degrees occur).
So instead of the Calabi-Yau threefolds $X^3_n$, we are now dealing
with the non-Calabi-Yau threefolds ${\cal B}^3_n=X^3_n/{\bf{Z}}_2$ which can
be written in Weierstrass form as follows:
\beqa
{\cal B}^3_n:\quad y^2=x^3+\sum_{k=-4}^4f_{4-{nk\over 2}}
(z_1)z_2^{4-k}x+\sum_{k=-6}^6g_{6-{nk\over 6}}(z_1)
z_2^{6-k}.\label{weierb}
\eeqa
The ${\cal B}^3_n$ are now elliptic fibrations over $F_{n/2}$ 
and still $K3$ fibrations over ${\bf{P}}^1_{z_1}$.
Note that the unmodded 3-folds
$X^3_n$ and the modded
spaces ${\cal B}^3_n$ have still the same $K3$-fibres with Picard number 
$\rho$.
The Euler numbers of ${\cal B}^3_n$ can be computed from the Euler numbers
of $X^3_n$ from the ramified covering as
\beqa \chi(X^3_n)=2\chi ({\cal B}^3_n)-2\cdot 24.\label{eulercalb}
\eeqa
Using $\chi(X^3_n)=2(1+\rho-h^{(2,1)}(X^3_n))$ and
$\chi({\cal B}^3_n)=2+2(1+\rho-h^{(2,1)}({\cal B}^3_n))$ we
derive the followong relation between $h^{(2,1)}(X^3_n)$ and $h^{(2,1)}({\cal
B}^3_n)$:
\beqa
h^{(2,1)}({\cal B}^3_n)={1\over 2}(\rho-2+h^{(2,1)}(X^3_n)-19).
\label{hrelation}
\eeqa
Second, $X^4_n$ are of course no more products ${\cal B}^3_n\times T^2$
but the torus $T^2_{z_4}$ now is a second elliptic fibre which varies
over ${\bf{P}}^1_{z_1}$. More precisely, this elliptic fibration just 
describes 
the emergence of
the del Pezzo  surface $dP_9$, which is given in Weierstrass form as
\beqa
dP_9:\quad y^2=x^3+f_4(z_1)x+g_6(z_1).\label{weierdp}
\eeqa
Therefore the spaces $X^4_n$ have the form
of being the following fibre products:
\beqa
X^4_n=dP_9\times_{{\bf{P}}^1_{z_1}}{\cal B}^3_n.\label{fibrep}
\eeqa
All $X^4_n$ are $K3$ fibrations
over the mentioned $dP_9$ surface.
The Euler numbers of all $X^4_n$'s are given by the value
\beqa
\chi=12\cdot 24=288.\label{euler}
\eeqa 
The corresponding (complex) three-dimensional IIB base manifolds $B^3_n$
have the following fibre product structure 
\beqa
B^3_n=dP_9\times_{{\bf{P}}^1_{z_1}}F_{n/2}.\label{fibrebase}
\eeqa
For the case already studied in \cite{CL}  with $n=0$, $B^3_0$
is just the product space $dP_9\times {\bf{P}}^1_{z_2}$.

The fibration structure of $X^4_n$ provides all necessary
information to compute the Hodge numbers of $X^4_n$ from the number
of complex deformations of ${\cal B}^3_n$, which we call $N_{{\cal B}^3_n}$.
These can be calculated from eq.(\ref{weierb}) and are summarized in table
(\ref{tab4}).
Note that in the cases $n> 2$ we have to subtract in the $N=2$ setup
$7+n-1=6+n$ reparametrizations, whereas in the $N=1$ setup only $6+n/2$
(for $n=0,2$ we have to subract 7 reparamerizations both for $N=1$ and $N=2$).

Knowing that the number of complex deformations of $dP_9$ is eight, as
easily be read in eq.(\ref{weierdp}), we
obtain for the number of complex structure deformations of $X^4_n$ the
following result:
\beqa
h^{(3,1)}(X^4_n)=8+3+N_{{\cal B}^3_n}=11+N_{{\cal B}^3_n}.\label{h31}
\eeqa
Next compute the number of K\"ahler parameters, $h^{(1,1)}(X^4_n)$, of $X^4_n$.
Since $h^{(1,1)}(dP_9)=10$ we  obtain the formula
\beqa
h^{(1,1)}(X^4_n)=10+\rho ,\label{h11}
\eeqa
where $\rho$ is the Picard number of the $K3$ fibre of $X^4_n$.

Finally, for the computation of $h^{(2,1)}(X^4_n)$ of $X^4_n$ we can use the
condition \cite{SVW} of tadpole cancellation, which tells us that
$h^{(1,1)}(X^4_n)-h^{(2,1)}(X^4_n)+h^{(3,1)}(X^4_n)={\chi\over 6}-8$.
Hence we get for $h^{(2,1)}(X^4_n)$
\beqa
h^{(2,1)}(X^4_n)=\rho+N_{{\cal B}^3_n}-19.\label{h21}
\eeqa
Using eqs.(\ref{h31},\ref{h11},\ref{h21}), we have summarized the 
spectrum of Hodge numbers
of $X^4_n$ in table (\ref{tab4}).
The computation of these 4-fold Hodge numbers, which was based on the counting
of complex deformations of the Weierstrass form eq.(\ref{weierb}), can be
checked in a rather independent way, by noting that $h^{(2,1)}(X^4_n)=
h^{(2,1)}({\cal B}^3_n)$,
since $N_{{\cal B}^3_n}=(19-\rho)+h^{(2,1)}({\cal B}^3_n)$. 
Then, using eq.(\ref{hrelation}) and
table (\ref{tab1}) the Hogde numbers $h^{(2,1)}(X^4_n)$ in table (\ref{tab4})
are immediately verified.

Let us compute the spectrum $C_F$ and $C_{het}$ for our chain of models using. 
The Hodge numbers of
$B^3_n$ eq.(\ref{fibrebase}) are universally given as
$h^{(1,1)}(B^3_n)=11$, $h^{(2,1)}(B^3_n)=0$.
Thus we obtain using eq.(\ref{h11}) that
\beqa
r(V)=h^{(1,1)}(X^3_n)-12=\rho -2;\label{v}
\eeqa
observe that the rank of the $N=1$ four-dimensional gauge groups
agrees with the rank of the six-dimensional gauge groups of the corresponding
$N=2$ parent models (see eq.(\ref{novector1})).\footnote{The Abelian
vector fields which arise in the $N=2$ situation from the $T^2$ compactification 
from six to four dimensions do not appear in the modded $N=1$ spectrum -- see
the discussion in the next chapter.}
Second, using eqs.(\ref{h31},\ref{h21}) we derive that
\beqa
C_F=38-r(V)+2h^{(2,1)}(X^4_n)=2+\rho+2N_{{\cal B}^3_n}.\label{c}
\eeqa
Using eq.(\ref{hrelation}), $C_F$ can be expressed by the number of 
hypermultiplets of the $N=2$ parent models as follows:
\beqa 
C_F=38+n_H-20.\label{ch}
\eeqa
This relation will become clear when considering the dual heterotic models
in the next chapter.
The explicit results for $N_{{\cal B}^3_n}$, $h^{(1,1)}(X^4_n)$, 
$h^{(2,1)}(X^4_n)$, $h^{(3,1)}(X^4_n)$, 
$r(V)$ and $C$ are contained in the
following table:

\beqa
%\begin{center}
\begin{tabular}{|c||c|c|c|c|c|c|} \hline
  $n$     & $N_{{\cal B}^3_n}$ & $h^{(1,1)}(X^4_n)$ & 
  $h^{(2,1)}(X^4_n)$ & $h^{(3,1)}(X^4_n)$ &
   $r(V)$  & $C$ \\ \hline\hline
0  & 129 & 12 & 112  & 140 & 0 & 262   \\ \hline
2 & 129 & 12  & 112  & 140 & 0 & 262   \\ \hline
4 &  141 & 16 & 128 & 152 & 4& 290 \\ \hline
6 & 165 & 18 & 154  & 176 & 6 & 340 \\ \hline
12 & 249 & 20 &  240 &  260 & 8 & 510 \\ \hline
\end{tabular}
%\end{center}
\label{tab4}
\eeqa

The equations (\ref{v}) and (\ref{c}) count the numbers of Abelian vector
fields and the number of neutral chiral moduli fields
of the four-dimensional F-theory compactification. 
%additional
%massless fields in general arise by adding various branes which may wrap
%around some of the non-trivial cycles of $X^4_n$. In fact, the existence
%of the branes is required by the cancellation of an $F$-theory anomaly
%which takes the value of $-\chi /24$. This anomaly can be either cancelled
%by $\chi /24$ 3-branes which fill the four-dimensional space time.
%Alternatively the anomaly can be also cancelled by non-trivial
%background gauge fields on the $F$-theory 7-branes.
Let us now discuss the emergence of $N=1$ non-Abelian gauge groups 
together with their matter contents. Namely, 
non-Abelian gauge groups arise by constructing  7-branes over which the
elliptic fibration has an ADE singularity
\cite{KV}. 
Specifically,
one has to consider
a (complex) two-dimensional space, which is a codimension one subspace of
the type IIB base $B^3$, over which the elliptic fibration has a singularity.
(In order to avoid adjoint matter, the space $S$ must satisfy $h^{(2,0)}(S)=
h^{(1,0)}(S)=0$.)
The world volume of the 7-branes is then given by $R^4\times S$; if
$n$ parallel 7-branes coincide, one gets for example an $SU(n)$ gauge
symmetry, i.e. the elliptic fibration acquires an $A_{n-1}$ singularity.
$N_F$ chiral massless
matter fields in the fundamental representation of the non-Abelian gauge
group can be geometrically engineered by bringing $N_f$ 3-branes
near the 7-branes, i.e close to $S$ \cite{BJPSV}.
The Higgs branches of these gauge theories should then be identified with
the moduli spaces of the gauge instantons on $S$. 
%So the pure Higgs branch
%corresponds to a situation where all 3-branes are replaced by the instantons,
%and the anomaly is entirely cancelled by  $\chi/24$ non-Abelian gauge
%instantons.
%The mixed branches are those with both 3-branes and instantons.
%We must emphasize that the formulas eqs.(\ref{v},\ref{c}) apply for the
%generic situation in moduli space, where the gauge group is Higgsed
%as much as possible.
%As we will show for our class of models this variety of Higgs branches 
%corresponds to the branches in the moduli spaces of bundles on the
%Calabi-Yau three-folds in the dual heterotic models.

%To be specific, since in our class of models the Euler number of all
%considered four-folds $X^4_n$ is $\chi=12\cdot 24=288$, the anomaly
%can be cancelled by 12 3-branes which fill space time.
In our class of models, 
the space $S$ is just given
by the
$dP_9$ surface which is the base of the $K3$ fibration of $X^4_n$.
The singularities of the elliptic four-fold fibrations are given by the
singularities of the Weierstrass curve for ${\cal B}^3_n$, given in 
eq.(\ref{weierb}). So the non-Abelian gauge groups arise at the
degeneration loci of eq.(\ref{weierb}). However, with this observation
we are in the same situation as in the $N=2$ parent models, since
the singularities of the modded elliptic curve ${\cal B}^3_n$ precisely agree
with the singularities of the elliptic 3-folds $X^3_n$ in eq.(\ref{weier}).
In other words, the non-Abelian gauge groups in the $N=2$ and $N=1$
models are identical. This observation can be explained from the fact
that the gauge group enhancement already occurs in eight dimensions
at the  degeneration loci of the elliptic $K3$ surfaces as F-theory
backgrounds. However, the underlying eight-dimensional $K3$ singularities
are not affected by the ${\bf{Z}}_2$ operation on
the coordinate $z_1$, but are the same in eqs.(\ref{weier}) and 
(\ref{weierb}).  

In the following section about the
heterotic dual models, we will explicitly determine the non-Abelian
gauge groups and the possible Higgsing chains. We will show
that after maximal Higgsing of the gauge groups the dimensions
of the instanton moduli spaces, being identical with the dimensions 
of the Higgsing moduli spaces, precisely agree with $2h^{(2,1)}(X^4_n)-\rho+2$
on the $F$-theory side; 
in addition
we also verify that the ranks of the  unbroken gauge groups after
the complete Higgsing precisely match the ranks of the F-theory 
gauge groups, as given by $r(V)$ in table eq.(\ref{tab4}).
 
\newpage 
%------------------------------------------------------------------------------
\subsection{Heterotic String on the $CY^{19,19}$}
%------------------------------------------------------------------------------

%------------------------------------------------------------------------------
\subsubsection {The $N=2$ models: the heterotic string on 
$K3(\times T^2)$}
%------------------------------------------------------------------------------
In this section, we will construct the  heterotic string compactifications
dual to $X^4_n$
with $N=1$ supersymmetry by ${\bf{Z}}_2$ modding of $N=2$ heterotic string 
compactications which are the duals of the F-theory models on $X^3_n\times 
T^2$. 
Heterotic string models with $N=2$ supersymmetry in four 
dimensions are obtained
by compactification on $K3\times T^2$ plus the specification of
an $E_8\times E_8$ gauge bundle over $K3$.
So in the heterotic context, we have to specify how the ${\bf{Z}}_2$ modding 
acts both on the compactification space $K3\times T^2$ as well
as on the heterotic gauge bundle.

As above, the $N=2$ heterotic models,  that are dual to the 
F-theory compactifications
on $X^3_n\times T^2$, are characterized by turning on $(n_1,n_2)=
(12+n,12-n)$ ($n\geq 0$)
instantons of the heterotic gauge group $E_8^{(I)}\times E_8^{(II)}$.
Recall further, while the 
first $E_8^{(I)}$ is generically completely broken by the
gauge instantons, the second $E_8^{(II)}$ 
is only completely broken for the cases
$n=0,1,2$; for bigger values of $n$ there is a terminating gauge
group $G^{(II)}$ of rank $r(V)$ which cannot be broken further.
The quaternionic dimensions of the instanton moduli space of $n$
instantons of a gauge group $H$, living on  $K3$,
was given by
$\dim_Q({\cal M}_{\rm inst}(H,k))=c_2(H)n-\dim H,\label{n2inst}$
where $c_2(H)$ is the dual Coxeter number of $H$.
Then, in the examples we are discussion, we derive the following formula
for the quaternionic dimension of the instanton moduli space:
\beqa
\dim_Q{\cal M}_{\rm inst}(E_8^{(I)}\times H^{(II)},n))=
112+30n+(12-n)c_2(H^{(II)})-\dim H^{(II)};
\label{hetinst}
\eeqa
here 
(for $n\neq 12$) $H^{(II)}$ is the 
commutant of the unbroken gauge group $G^{(II)}$ in 
$E_8^{(II)}$. Specifically, the following gauge groups $G^{(II)}$ and
dimensions of instanton moduli spaces are derived:
\beqa
%\begin{center}
\begin{tabular}{|c||c|c|} \hline
  $n$     & $G^{(II)}$  & $\dim_Q{\cal M}_{\rm inst}$ \\ \hline\hline
0  &   1 & 224   \\ \hline
2 &   1 & 224   \\ \hline
4 &    $SO(8)$& 252 \\ \hline
6 & $E_6$ & 302 \\ \hline
12 & $E_8$ & 472 \\ \hline
\end{tabular}
%\end{center}
\label{tab5}
\eeqa
The number of massless gauge singlet hypermultiplets is then simply given by
\beqa
n_H=20+\dim_Q{\cal M}_{\rm inst},\label{heth}
\eeqa
where 20 corresponds to the complex deformations of $K3$.
One finds perfect agreement, if one
compares the spectra of F-theory on the 3-folds $X^3_n$
(see eqs.(\ref{nohyper},\ref{novector1}) and table (\ref{tab1})) with
the spectra of the heterotic string on $K3$ with instanton numbers 
$(12+n,12-n)$(see eq.(\ref{heth}) and table (\ref{tab5})).
\\
{\bf{Note:}} On the heterotic side there is an perturbative gauge symmetry
enhancement at special loci in the hypermultiplet moduli spaces.
Specifically, by embedding the $SU(2)$ holonomy group of
$K3$, namely the $SU(2)$ bundles with instanton numbers
$(12+n,12-n)$ in $E_8^{(I)}\times E_8^{(II)}$, the six-dimensional gauge group
is broken to $E_7^{(I)}\times E_7^{(II)}$ (or $E_7^{(I)}\times 
E_8^{(II)}$ for $n=12$); 
in addition one gets charged hyper multiplet fields, which can be used to
Higgs the gauge group via several intermediate gauge groups down to
the terminating groups. 
The dimensions of the Higgs moduli space,
i.e. the number of gauge neutral hypermultiplets, agrees with the dimensions
of the instanton moduli spaces eq.(\ref{hetinst}).

%------------------------------------------------------------------------------
\subsubsection {The $N=1$ models: the heterotic string on $Z=(K3\times
T^2)/{\bf{Z}}_2$}
%------------------------------------------------------------------------------
Now let us construct the four-dimensional heterotic compactifications
with $N=1$ supersymmetry, which are dual to F-theory on $X^4_n$,
by ${\bf{Z}}_2$ modding of the heterotic string compactications 
on $K3\times T^2$.
In the first step we discuss the ${\bf{Z}}_2$ modding of 
the compactication space $K3\times T^2$ which
results in a particular Calabi-Yau 3-fold $Z$:\footnote{At the orbifold point
of $K3$ one can construct $Z$ as $T^6/({\bf{Z}}_2\times {\bf{Z}}_2)$, where 
one of the
${\bf{Z}}_2$'s acts freely, see e.g. \cite{CL}.}
\beqa
Z={(K3\times T^2)\over {\bf{Z}}_2}.\label{hetmodding}
\eeqa
Specifically, the ${\bf{Z}}_2$-modding reduces $K3$ to the 
del Pezzo surface $dP_9$. 
This corresponds to having on K3 a Nikulin involution of type
(10,8,0) with two fixed elliptic fibers 
in the K3 leading to
\beqa
\begin{array}{ccc}K3&\rightarrow &dP_9\\\downarrow & &\downarrow\\{\bf P}^1_y&
\rightarrow &{\bf P}^1_{\tilde{y}}\end{array}
\eeqa
induced from the quadratic base map $y\rightarrow \tilde{y}:=y^2$ with the 
two branch points $0$ and $\infty$ (being the identity along the fibers).
In the Weierstrass representation of $K3$
\beqa
K3:\quad y^2=x^3-f_8(z)x-g_{12}(z),\label{weierk3}
\eeqa
the mentioned quadratic redefinition translates to the representation 
\beqa
dP_9:\quad y^2=x^3-f_4(z)x-g_6(z)\label{weierdpa}
\eeqa
of $dP_9$ (showing again the $8=5+7-3-1$ deformations).
So the ${\bf{Z}}_2$ reduction of $K3$ to the
non Ricci-flat $dP_9$ corresponds to the
reduction of the $X^3_n$ to the non Calabi-Yau space ${\cal B}^3_n$
(cf. eqs.(\ref{weier} and \ref{weierb})).
Representing $K3$ as a 
complete intersection in the product of projective spaces
as $K3={\scriptsize \left[\begin{array}{c|c}{\bf P}^2&3\\{\bf P}^1&2
\end{array}\right]}$,
the ${\bf{Z}}_2$ modding reduces 
the degree in the ${\bf P}^1$ variable by half; hence the
$dP_9$ can be represented as
$dP_9={\scriptsize \left[\begin{array}{c|c}{\bf P}^2_x&3\\{\bf{P}}^1_y&1
\end{array}
\right]}$. 
This makes 
visible on the one hand its elliptic fibration over ${\bf{P}}^1$
via the projection onto the second factor; on the other hand the defining
equation $C(x_0,x_1,x_2)y_0+C^\prime (x_0,x_1,x_2)y_1=0$ shows that the 
projection onto the first factor exhibits $dP_9$ as being a ${\bf P}^2_x$ 
blown up in
9 points (of $C\cap C^\prime$), thus having as nontrivial hodge number (besides
$b_0,b_4$) only $h^{1,1}=1+9$. 
Furthermore, the $dP_9$ has 8 complex structure moduli: they can be seen
as the parameter input in the construction of 
blowing up the plane in the 9 intersection points of two cubics
(the ninth of which is then always already determined as they sum up to
zero in the addition law on the elliptic curve;
so one ends up with $8\times 2-8$ parameters). 
 
As in the dual F-theory description a second $dP_9$ emerges by fibering
the $T^2$ in eq.(\ref{hetmodding}) over the ${\bf P}^1$ base of $dP_9$.
So in analogy to eq.(\ref{fibrep}) the heterotic Calabi-Yau 3-fold
$X^3_{\rm het}$, which is elliptically
fibered over $dP_9$, has the following fiber product structure
\beqa
Z=dP_9\times_{\bf P^1} dP_9.\label{hetfibrep}
\eeqa
The number of K\"ahler  deformations of $Z$ is given by the sum of
the deformations of the two $dP_9$'s minus one of the common ${\bf P}^1$ base,
i.e. $h^{(1,1)}(Z)=19$. Similarly we obtain
$h^{(2,1)}(Z)=8+8+3=19$.
This Calabi-Yau 3-fold is in fact well known, being one of the Voisin-Borcea
Calabi-Yau spaces. It can be obtained from $K3\times T^2$ 
by the Voisin-Borcea involution, which consists in the `del Pezzo' involution
(type (10,8,0) in Nikulins classification) with two fixed elliptic fibers 
in the K3 combined with the 
usual ``-"-involution with four fixed points in the $T^2$. 
Writing $K3\times T^2$ as $K3\times T^2={\tiny 
\left[\begin{array}{c|cc}{\bf P}^2&3&0\\{\bf P}^1&0&2\\{\bf P}^2&0&3
\end{array}\right]}$
the Voisin-Borcea involution changes this to
$Z={\tiny 
\left[\begin{array}{c|cc}{\bf P}^2&3&0\\{\bf P}^1&1&1\\{\bf P}^2&0&3
\end{array}\right]}=
dP\times _{{\bf P}^1}dP$.
%So here the symmetric degree one entries in the 
%$P^1$ variables have a seemingly different origin: one by {\it `reduction'} 
%(from 
%two) and one by {\it `emergence'} (from zero).
Observe that the base of the elliptic fibration of $Z$ is given
by the $dP_9$ surface which {\it `emerges'} 
(from the trivial elliptic fibration) after the ${\bf{Z}}_2$ modding.

After having described the ${\bf{Z}}_2$ modding of $K3\times T^2$, we will now
discuss how this operation acts on the heterotic gauge bundle.
Recall that in the $N=2$ heterotic models on $K3\times T^2$ the heterotic
gauge group $E_8^{(I)}\times E_8^{(II)}$ lives on the four-dimensional
space $K3$. We will now consider a $N=1$ situation where, 
after the ${\bf{Z}}_2$
modding, the heterotic gauge group still lives on a four manifold, namely on
the del Pezzo surface $dP_9$, which arises from the ${\bf{Z}}_2$ modding of
the $K3$ surface.
Then the complex dimension of the instanton moduli space of $k$
gauge instantons of a gauge group $H$, which lives on $dP$, is given by
\beqa
\dim_C {\cal M}_{\rm inst}(H,k)=2c_2(H)k-\dim H.\label{complexdim}
\eeqa

The action of the ${\bf{Z}}_2$ modding on the gauge bundle is now defined 
in such a way that
the gauge instanton numbers are reduced by half (think of the limit case of
pointlike instantons):
\beqa
k_{1,2}={n_{1,2}\over 2}.\label{nkhalfe}
\eeqa
So the total number of gauge instantons in $E_8^{(I)}
\times E_8^{(II)}$ will be reduced by two, i.e. $k_1+k_2=12$ and we 
consider $(k_1,k_2)=
(6+{n\over 2},6-{n\over 2})$ instantons in $E_8^{(I)}\times E_8^{(II)}$.
The reduction of the total instanton number by half from 24 to 12
can be explained from the observation that on the F-theory side
the tad-pole anomaly can canceled either by $\chi/24=12$
3-branes  or by 12
gauge instantons of the  gauge group $H$, which lives over the
four manifold $S=dP_9$.
%Therefore one needs either 12 5-branes in the heterotic dual models
%\cite{FMW} or the background of 12 gauge instantons. 
So, with $k_1+k_2=12$ and using eq.(\ref{complexdim}),
we can compute the complex dimensions of the instanton moduli space for the
gauge group $E_8^{(I)}\times H^{(II)}$, where again $H^{(II)}$ is the commutant
of the gauge group $G^{(II)}$ which cannot be further broken by the
instantons:
\beqa
\dim_C{\cal M}_{\rm inst}(E_8^{(I)}\times H^{(II)},n))=
112+30n+(12-n)c_2(H^{(II)})-\dim H^{(II)}.\label{hetinsta}
\eeqa
 This result precisely agrees with the quaternionic dimensions of the
instanton moduli space, eq.(\ref{hetinst}), in the unmodded $N=2$ models.
So we see that we obtain 
as gauge bundle deformation parameters of the heterotic string on 
$Z$
the same number of massless, gauge neutral $N=1$
{\it chiral} multiplets as the number of massless $N=2$ {\it hyper} multiplets
of the heterotic string on $K3$.
This means that the ${\bf{Z}}_2$ modding keeps 
just one of the two chiral fields
in each $N=2$ hyper multiplet in the massless sector.
These chiral multiplets describe the Higgs phase of the $N=1$ heterotic
string compactifications.

The gauge fields in $N=1$ heterotic string compactifications on $Z$
are just given by those gauge fields which arise from the compactification
of the heterotic string on $K3$ to six dimensions; therefore they are
invariant under the ${\bf{Z}}_2$ modding. However, the complex scalar fields
of the corresponding $N=2$ vector multiplets in four dimensions
do not survive the ${\bf{Z}}_2$ modding. Therefore, there is no Coulomb phase
in the $N=1$ models in contrast to the  $N=2$ parent compactifications.
Also observe that the two vector fields, commonly denoted by $T$ and $U$,
which arise from the compactification from six to four dimensions on $T^2$,
disappear from the massless spectrum
after the modding. This is expected since the
Calabi-Yau space has no isometries which can lead to massless gauge
bosons. Finally, the $N=2$ dilaton vector multiplet $S$ is reduced
to a chiral multiplet in the $N=1$ context.

These relations between the spectra of the $N=1$ and $N=2$ models can
be understood from the observation that the considered ${\bf Z}_2$ modding
corresponds to a spontaneous breaking of $N=2$ to $N=1$ spacetime
supersymmetry \cite{KK}.

In summary, turning on $(6+{n\over 2},6-{n\over 2})$
gauge instantons of $E_8^{(I)}\times E_8^{(II)}$
in our class of $N=1$ heterotic string compactifications on
$Z$, the unbroken gauge groups $G^{II}$ as well as the number
of remaining massless chiral fields 
(not counting the geometric moduli from $Z$, see next paragraph) 
agree with the unbroken gauge groups
and the number of massless hyper multiplets (again without
the 20 moduli from $K3$) in the  heterotic 
models on $K3$ 
with $(12+n,12-n)$ gauge instantons.
The specific gauge groups and the numbers of chiral fields are 
already summarized in table (\ref{tab5}).

Now comparing with the F-theory spectra, we first observe
that the ranks of the gauge groups after
maximally possible Higgsing perfectly match in the two dual
descriptions (see tables (\ref{tab4}) and (\ref{tab5})).

Next compare the number  of chiral $N=1$  
moduli fields in the heterotic/F-theory
dual pairs. 
First, looking at the Hodge numbers of the dual F-theory fourfolds
$X^4_n$, as given in table (\ref{tab4}) we recognize that
\beqa 
2h^{(2,1)}(X^4_n)-(\rho -2)=\dim_C {\cal M}_{\rm inst}.\label{agree}
\eeqa

Let us argue this independent of the case by case calculation. 
Namely, using $C_F=38+(2h^{(2,1)}(X^4_n)-(\rho-2))=n_H+18$ 
(cfr. eqs.(\ref{c}) and
(\ref{ch})) and $n_H=20+\dim_Q{\cal M}_{\rm inst}=
20+\dim_C{\cal M}_{\rm inst}$,
%$e_{CY^3}=2\cdot e_{{\cal B}}-2\cdot 24$ by the ramified covering
%and $e_{CY^3}=2(1+\rho -h^{2,1}(CY^3))$,
%$e_{{\cal B}}=2+2(1+\rho -h^{2,1}({\cal B}))$ one gets the result
%from the number of hypermultiplets 
%$\sharp H=h^{2,1}(CY^3)+1=20+\dim_Q {\cal M}_{inst}$ 
%and using that 
%$h^{2,1}(X^4)=h^{2,1}({\cal B})$ as
%$N_{{\cal B}^3_n}=(19-\rho)+h^{2,1}({\cal B})\leftrightarrow
%h^{3,1}=8+3+N_{{\cal B}^3_n}=30-\rho +h^{2,1}({\cal B})\leftrightarrow
%h^{2,1}=h^{1,1}+h^{3,1}-40=h^{2,1}({\cal B})$.
%
%SHOW $N_{{\cal B}^3_n}=(19-\rho)+h^{2,1}({\cal B})$.
%
one gets 
\beqa
C_F=38+(2h^{(2,1)}(X^4_n)-(\rho -2))=38+\dim_C {\cal M}_{\rm inst}
\eeqa

On the 
heterotic side, the total number $C_{het}$ of chiral moduli fields is 
given by the dimension of the gauge
instanton moduli space plus the number of geometrical moduli 
$h^{(1,1)}(Z)+h^{(2,1)}(Z)$
from the
underlying Calabi-Yau space $Z$, 
which is 38 for our class of models, i.e.
\beqa
C_{het}=38+\dim_C {\cal M}_{\rm inst}.\label{chet}
\eeqa
\\
{\bf{Discussion:}} So we have shown that the massless spectra
of Abelian vector multiplets and of the gauge singlet chiral plus
antichiral fields agree for all considered dual pairs.
The next step in the verification of the $N=1$ string-string duality
after the comparison of the massless states
is to show that the interactions, i.e. the $N=1$ effective action,
agree. In particular, the construction of the superpotentials
is important to find out the ground states of these theories.
This was already done \cite{DGW,CL} for one particular model 
(the model with $k=9$, $n=0$), where on the heterotic,
side the superpotential was entirely generated by world sheet instantons.
It would be interesting to see whether space time instantons
would also contribute to the heterotic superpotential in some other models 
and whether supersymmetry can be broken by the
superpotential. In addition, it would also be interesting to compare
the holomorphic gauge kinetic functions in $N=1$ dual string pairs, in
particular in those models obtained from $N=2$ dual pairs
by ${\bf{Z}}_2$ moddings respectively by spontaneous supersymmetry
breaking from $N=2$ to $N=1$.
%------------------------------------------------------------------------------
\subsubsection{Non-Abelian gauge groups and Higgsing chains}
%------------------------------------------------------------------------------
For the computation of $C$ and $r(V)$ we have considered  a generic
point in the moduli space where the gauge group is broken as far as possible
to the group $G^{II}$
by the vacuum expectation values of the chiral fields.
In this section, we now want to determine the non-Abelian gauge groups
plus their matter content which arise in special loci of the
moduli space. 
Since the $N=1$ gauge bundle is identical to the $N=2$ bundle, which is given
by $SU(2)\times SU(2)$, the maximally unbroken gauge group is
for $k_1,k_2\geq 3$ (i.e. $n\leq 6$)
given by $E_7^{(I)}\times E_7^{(II)}$;
the $N=1$ chiral representations follow immediately from the
$N=2$ hypermultiplet representations and transform as

\beqa
E_7\times E_7:\quad (k_1-2) ({\bf 56},{\bf 1})  
 +
(k_2-2) ({\bf 1},{\bf 56})
+ (4(k_1+k_2)-6)({\bf 1},{\bf 1}).
\label{e7}
\eeqa
Higgsing $E_7^{(I)}\times E_7^{(II)}$ to $E_6^{(I)}\times E_6^{(II)}$
 one is left with  
chiral matter fields in the following representations of the
gauge group $E_6^{(I)}\times E_6^{(II)}$:
\beqa
E_6\times E_6:\quad (k_1-3)\lbrack ({\bf 27},{\bf 1}) + ({\bf\overline{27}},
{\bf 1})\rbrack &+&
(k_2-3)\lbrack ({\bf 1},{\bf 27})
+({\bf 1},{\bf\overline{27}})\rbrack \nonumber\\ 
+ (6(k_1+k_2)-16)({\bf 1},{\bf 1}).
\label{e6}
\eeqa
When $k_1=12$ ($n=12$) the gauge group is $E_6^{(I)}\times E_8^{(II)}$
with chiral matter fields
\beqa
E_6\times E_8:\quad 
9\lbrack ({\bf 27},{\bf 1}) + ({\bf\overline{27}},{\bf 1})\rbrack 
+ 64({\bf 1},{\bf 1}).\label{e6e8}
\eeqa
Since $k_1\geq 6$, the number of $({\bf 27},{\bf 1}) + 
({\bf\overline{27}},{\bf 1})$ is always big enough that the first
$E_6^{(I)}$ can be completely broken. On the other hand, only
for the cases $n=0,2$ the group $E_6^{(II)}$ can be completely Higgsed away
by giving vacuum expectation values to the fields
$({\bf 1},{\bf 27})
+({\bf 1},{\bf\overline{27}})$.
For $k_2=4$ ($n=4$), $E_6^{(II)}$ can be only Higgsed to the group 
$G^{(II)}=SO(8)$, and
for $k_2=3$ ($n=6$) 
there are no charged fields
with respect to $E_6^{(II)}$ such that the terminating gauge group is just 
$E_6^{(II)}$.
Clearly, the ranks of these
gauge groups are in agreement with the previous discussions, i.e.
with the results for $r(V)$; in addition,
assuming maximally possible Higgsing of both gauge group factors the complex
dimension
of the Higgs moduli space agrees with the dimensions of the instanton
moduli spaces as given in eq.(\ref{hetinsta}) and in table (\ref{tab5}).

Consider the Higgsing of, say, the first gauge group $E_6^{(I)}$.
Namely,
like in the
$N=2$ cases \cite{CF}, \cite{AFIQ}, it 
can be Higgsed through the following chain of Non-Abelian
gauge groups:
\beqa
E_6\rightarrow SO(10)\rightarrow SU(5)\rightarrow SU(4)\rightarrow
SU(3)\rightarrow SU(2)\rightarrow SU(1).\label{chain}
\eeqa
In the following we list the spectra for all gauge groups within this chain:
\beqa
SO(10)& :&
\quad
(k_1-3)( {\bf 10} + {\bf\overline{10}})  +
(k_1-4)({\bf 16}+{\bf\overline{16}})+(8k_1-15){\bf 1}
,\label{so10}\\
%\eeqa
%\beqa
SU(5)& :&
\quad
(3k_1-10)( {\bf 5} + {\bf\overline{5}})  +
(k_1-5)({\bf 10}+{\bf\overline{10}})+(10k_1-24){\bf 1}
,\label{su5}\\
%\eeqa
%\beqa
SU(4)& :&
\quad
(4k_1-16)( {\bf 4} + {\bf\overline{4}})  +
(k_1-5)({\bf 6}+{\bf\overline{6}})+(16k_1-45){\bf 1}
,\label{su4}\\
%\eeqa
%\beqa
SU(3)& :&
\quad
(6k_1-27)( {\bf 3} + {\bf\overline{3}})  +(24k_1-78){\bf 1}
,\label{su3}\\
%\eeqa
%\beqa
SU(2)& :&
\quad
(12k_1-56) {\bf 2} +  (36k_1-133){\bf 1}
,\label{su2}\\
SU(1)& : & \quad (60 k_1-248){\bf 1}.\label{su1}
\eeqa
In order to keep  contact with our previous discussion, we see that
the number of massless chiral fields at a generic point in the
moduli space, i.e. for complete Higgsing down to $SU(1)$, is given
by $60 k_1-248 +2k_2c_2(H^{(II)})-\dim H^{(II)}$ which precisely
agrees with $\dim_C{\cal M}_{\rm inst}$
\newpage
%------------------------------------------------------------------------------
\subsection{Standard embedding and Higgsing}
%------------------------------------------------------------------------------
In this final section, let us consider the standard embedding \cite{A2}!
We will do this for two reasons: first, the standard 
embedding involves the anomaly cancellation without fivebranes ($n_5=0$).
Thus, under duality we should expect that the number of threebranes 
vanishes also.
Then assuming no four-flux is turned on and also no instantons, inside the 
compact part of the worldvolume of the seven-brane, are tuned on, we should 
expect $\chi(X_4)=0$; second, we can easily determine the number of 
matter multiplets transforming as ${\bf 27}$'s of $E_6$, since 
left handed ${\bf{27}}$'s come from elements of $H^1(Z,TZ)$, while 
right handed ${\bf\overline{27}}$'s come from elements of $H^1(Z,TZ^*)$ and 
the net amount of chiral matter is then given by (c.f. section 2)
$N_{gen}=\frac{1}{2}c_3(TZ)$. In addition we can determine the number
of bundle moduli $m_{bun}$, and we can try to understand the analog 
of the 6D complete Higgsing process!     
\\ 
Now, take the tangent bundle $TZ$ of $Z$ 
which has $SU(3)$ holonomy and identify
it with the gauge field that belongs to the subgroup of $E_8$ (which commutes
with $E_6$) and thus breaking $E_8$ to $E_6$. So one ends up with the gauge
group $E_8\times E_6$ where we think of the $E_8$ as being the "hidden
sector" and the $E_6$ as being the "observable sector" which leads to 
massless charged matter. The adjoint of $E_8$ decomposes under 
$E_6\times SU(3)$ as
\beqa
\quad
{\bf{248}}=({\bf{78,1}})+({\bf{27,3}})+({\bf\overline{27}},{\bf\overline 
{3}})+({\bf {1,8}}).  
\quad
\eeqa
The compactification is specified by $h^{1,1}(Z)$, $h^{2,1}(Z)$ and 
$h^1(Z, {\rm{End}}(TZ))$ where $TZ$ denotes the tangent bundle 
to $Z$. Further, let 
us assume that $Z$ can be represented by its smooth Weierstrass model. 
So we have (as above) $h^{1,1}(Z)=11-c_1^2$ and $h^{2,1}(Z)=11+29c_1^2$. 
We computed the index of $TZ$ in section 3.3.2, which we will use now, recall
that the index is given by $I=-46-28c_1^2$ and thus leading to the number of
bundle moduli
\beqa
m_{bun}=-46-28c_1^2+2n_o
\eeqa
Since $m_{bun}> 0$ we get a condition for the bundle moduli which are 
odd under the $\tau$-involution of $Z$: $n_o\geq 46+28c_1^2$. 
\\
{\bf{3-branes:}} Recall, the number of threebranes can be written as (c.f. 
section 5)
\beqa
\frac{\chi(X_4)}{24}=2+\frac{1}{4}(40+28c_1^2+I-16+r(V))=n_3
\eeqa
where $r(V)$ denotes the rank of the unbroken 4D gauge group. In our case 
we find with $I=-46-28c_1^2$ and $r(V)=14$ 
\beqa
n_3=0.
\eeqa
\\ \\
{\bf{Higgsing:}} In the standard embedding, the number of ${\bf{27}}$'s resp. 
${\overline{\bf{27}}}$'s is given by $h^{1,1}(Z)$ resp. $h^{2,1}(Z)$, so
we expect a total number of $22+28c_1^2$ charged multiplets.  
Let us now Higgs the $E_6$ completely, we get 
\beqa
C_{Higgs}=(22+28c_1^2)27-78=516+756c_1^2.
\eeqa
We are left with an unbroken $E_8$ and a number $C_{Higgs}$ of $N=1$
neutral chiral multiplets. Also taking into account the geometrical 
moduli ($h^{1,1}, h^{2,1}$) and the bundle moduli $m_{bun}$ we end up
with 
\beqa
C_{het}=492+756c_1^2+2n_o
\eeqa
further, we have the net generation number $N_{gen}=\frac{1}{2}c_3(V)=30c_1^2$.
\\
Let us try to shed some light on $C_{het}$! 

A striking fact is the appearance of $492$, indicating a possible relation
to the six-dimensional heterotic $(n_1,n_2)=(0,24)$ compactification which 
leads to an unbroken $E_8\times E_7$ gauge group (the standard embedding 
mentioned in section 5.2.1). In particular, complete Higgsing 
leads to $n_V=248$, $n_T=1$ and $n_H=427+45+20=492$. 
Now, following the 'spirit' of the last section, we should compactify
the 6D theory on $T^2$ leaving us with $N=2$ supersymmetry in 4D. 
To obtain a $N=1$ theory we have to mod out by a group $G$. 
To determine $G$ let us make a short digression! If we recall $Z$, we get
a hint from the Hodge numbers of $Z$ for a viable $G$. We find that for 
$c_1(B_2)=0$, we obtain the well known Voisin-Borcea model (11,11), which 
can be obtained from $K3\times T^2$ by the Nikulin involution of type 
(10,10,0) which has no fixed fibers in the $K3$, so we have 
$Z=\frac{K3\times T^2}{(\sigma,-1)}$ where '-1' is a involution on $T^2$ 
with four fixed points. Moreover, in this case we can think of $Z$ as being
elliptically 
fibered over the Enriques surface $K3/\sigma$. So, in the following
let us restrict ourselfs to the (11,11) model! Thus $C_{het}$ reduces to
$C_{het}=492+2n_o$. Let us return to our model on $K3\times T^2$. Now 
recall, the 4D $N=2$ spectrum is obtained from 6D, upon $T^2$ 
compactification, as follows: the 6D vectors become 4D vectors, the 
6D tensor becomes a abelian $N=2$ vector field, the hyper multiplets
become 4D $N=2$ hyper multiplets and in addition we get 2 vectors 
related to the $T$ and $U$ moduli of $T^2$. Let us discuss the modding
of $K3\times T^2$. We will consider that, after modding, the heterotic 
gauge group still lives on a four manifold, which is the Enriques surface.\
The complex dimension of the instanton moduli space is given by 
$\dim_C{\M}_{inst}=2nc_2(V)-n^2+1$ and if we assume that the modding
reduces the instanton numbers by half, we get $\dim_C{\M}_{inst}=45$.
So, we are left with the explanation of $2n_o$. Understanding $2n_o$ would 
then lead to an interesting picture, where the complete Higgsed 6D (0,24)
model would correspond to the 4D ($h^{1,1}, h^{2,1}=11,11$) model, where 
we consider again the complete Higgsed case of the 4D standard embedding.
We will close with this neat observation, which leads us to summary and 
outlook!
\newpage

%------------------------------------------------------------------------------
\addcontentsline{toc}{section}{Summary and Outlook}
\section*{Summary and Outlook}
%------------------------------------------------------------------------------

We have studied four-dimensional $N=1$ het/F-theory duality. 
The duality involves the understanding 
of vector bundles on elliptic fibrations. Vector bundles can be constructed 
using the parabolic or spectral cover bundles construction. We adopted the 
parabolic approach and constructed a class of $SU(n)$ vector bundles which
have $n$ odd and a certain congruence relation 
($\eta\equiv 0({\rm{mod}}\ \ n)$). 
In particular, these bundles have non-vanishing $c_3(V)$ (which is related 
to a non-zero net amount of chiral matter). Then we compared our results 
with the spectral cover approach and found agreement for a certain twist 
($\lambda=\frac{1}{2n}$) of the line bundle on the spectral cover. 

We also computed the number of bundle moduli of $E_8$ bundles, 
using a character valued index theorem, which was first used in its specific 
form by Friedman, Morgan and Witten for $SU(n)$ vector bundles.

Further, consistent F-theory compactification requires a number of space-time
filling threebranes, which should turn, under duality, 
into heterotic fivebranes, which 
wrapping the elliptic fiber. This 
matching has been established by Friedman, Morgan and 
Witten for $E_8$ bundle, leaving no unbroken gauge group. We could extend 
this matching to the general case
of $SU(n)$ vector bundles, leaving an unbroken gauge group on the heterotic
side. 

A het/F-theory duality also involves the comparision between massless 
spectra. We found a complete matching of the spectra under two assumptions: 
first, no four-flux is turned on; second, no instanton background inside the
compact part of the sevenbrane is turned on.   
      
Then we considered $N=1$ dual het/F-theory pairs. In particular, we 
constructed a chain of dual pairs by ${\bf{Z}}_2$-modding of corresponding 
dual pairs with $N=2$ supersymmetry, which were obtained from six-dimensional
dual pairs after $T^2$ compactification. 

Apart from the moduli matching, one should also 
understand the part of the 4D
heterotic spectrum which corresponds to charged matter. We made a first step 
in this direction due to the computation of $c_3(V)$ in the parabolic
approach; recently, this computation was also performed in the spectral 
cover approach \cite{CP}. However, in both approaches a difficulty appears.
For $\tau$-invariant vector bundles, we can apply an index theorem to 
compute the number of bundle moduli, but these bundles have $c_3(V)=0$ and 
therefore lead to a zero net amount of chiral matter; 
in contrast, vector bundles with $c_3(V)\neq 0$ are not 
$\tau$-invariant and thus we lose the
control over the bundle moduli (since we cannot apply the character valued
index theorem). This seems to be a rather technical problem, but it has to be 
solved in order to understand possible couplings between gauge singlets 
and chiral matter fields. 

Another problem which begs further analysis is to work out a refined 
het/F-theory matter dictionary, which requires an improved understanding of 
intersecting sevenbranes \cite{BJPS}.

\newpage

%-----------------------------------------------------------------------------
%\section{Bibliography}
\addcontentsline{toc}{section}{References}
%\bibliography{3}
%\bibliographystyle{utphys}
%\begin{thebibliography}
\begingroup\raggedright\endgroup

\end{document}